\documentclass[sigconf,nonacm]{acmart}
\setcopyright{cc}
\setcctype[4.0]{by}

\copyrightyear{2026}
\acmYear{2026}

\usepackage{todonotes}
\usepackage{wasysym}
\usepackage{subcaption}
\usepackage{xspace}
\newcommand\aisandbox{\textit{AI Sandbox}\xspace}
\newcommand\mlguide{\textit{mlguide}\xspace}
\newcommand\mlkitchen{\textit{mlkitchen}\xspace}
\newcommand\maratus{\textit{maratus}\xspace}
\newcommand\mlflow{\textit{mlflow}\xspace}
\newcommand\colab{\textit{Google Colab}\xspace}

\begin{document}

\title[Designing Accessible and Rigorous ML Experiments]
{Designing Accessible and Rigorous ML Experiments: No-Code Visual Workflows and Structured Guidance for Beginners}

\def\mevis{
    \affiliation{%
      \institution{Fraunhofer Institute for Digital Medicine MEVIS}
      \streetaddress{Max-von-Laue-Str. 2}
      \city{Bremen}
      \country{Germany}
      \postcode{28359}
    }
}

\author{Henrik Detjen}
\email{henrik.detjen@mevis.fraunhofer.de}
\orcid{0000-0002-7683-1797}
\mevis{}

\author{David Pfrang}
\email{david.pfrang@mevis.fraunhofer.de}
\orcid{0000-0000-0000-0000}
\mevis{}

\author{Max Westphal}
\email{max.westphalg@mevis.fraunhofer.de}
\orcid{0000-0000-0000-0000}
\mevis{}



\author{Melissa Finster}
\email{melissa.finster@mevis.fraunhofer.de}
\orcid{0000-0000-0000-0000}
\mevis{}



\author{Dörte Corr}
\email{doerte.corr@mevis.fraunhofer.de}
\orcid{0000-0000-0000-0000}
\mevis{}

\author{Markus Wenzel}
\email{markus.wenzel@mevis.fraunhofer.de}
\orcid{0000-0000-0000-0000}
\mevis{}

\renewcommand{\shortauthors}{Detjen, et al.}
\begin{abstract}
    Designing machine learning (ML) experiments requires beginners to plan a methodologically sound workflow and implement it in a concrete environment. We evaluated two forms of support in two within-participant studies with 18 predominantly technically trained participants: \mlguide, which provides structured, evidence-backed guidance for workflow design, and \aisandbox, a visual no-code environment for workflow implementation. After using \mlguide, participants produced better high-level workflow designs, but their scores on a quiz about specific methodological choices did not improve significantly. Compared with Google Colab, \aisandbox received higher hedonic user experience ratings, while perceived workload and acceptance did not differ significantly. These findings suggest that different forms of interactive support help with different aspects of beginner ML experimentation. The paper contributes a reusable evaluation task with instructions, a modified dataset, and a quiz, as well as empirical evidence on structured guidance for workflow design and on the experience of implementing workflows in visual and notebook-based environments.

\end{abstract}



\maketitle


\section{Introduction}

Machine-learning (ML) experimentation has become part of everyday work in research, engineering, and applied domains well beyond specialist ML teams. Yet designing a sound ML experiment remains difficult, especially for beginners and occasional practitioners. The difficulty is not only technical. Users must decide \emph{what} a sensible workflow should contain---for example, how data preparation, model training, validation, comparison, and reporting should relate to one another---and also \emph{how} that workflow should be realised in a concrete tool. Recent work on ML-based science has underlined the importance of these choices: recurring problems of validity, reproducibility, and generalisability often arise not from model fitting alone, but from weaknesses in the surrounding experimental workflow and reporting practice \cite{2024_Kapoor_REFORMSConsensusbasedRecommendations}. This makes beginner support an interactive-systems problem as much as an algorithmic one.

A useful way to frame this challenge is to distinguish between two coupled layers of understanding. First, users need a \emph{conceptual model} of an ML workflow: what the main stages are, how they should be ordered, and what kinds of choices each stage demands. Second, users need an \emph{implementation model}: how those stages are instantiated in a programming environment, notebook, or visual workflow editor. These two layers are related, but prior work suggests that they are not interchangeable. In interactive ML, user-interface design has long been recognised as central to whether non-experts can contribute effectively, inspect system behaviour, and remain in control of the learning process \cite{2018_Dudley_ReviewUserInterface}. However, different tool families support different layers of the problem.

Computational notebooks such as Jupyter are widely used because they combine executable code, narrative text, and outputs in a single environment \cite{kluyver2016jupyter}. At the same time, HCI studies show that notebook work is shaped by tension between exploration and explanation, by difficulties in tracking variants over time, and by recurring pain points around structure, navigation, and reuse \cite{rule2018notebooks,kery2018story,chattopadhyay_whats_2020}. Visual workflow systems such as KNIME and Orange make pipeline structure more explicit by representing analysis steps as composable graphs of operations rather than free-form code \cite{berthold2008knime,demsar2013orange}. Educational systems such as Teachable Machine and TensorFlow Playground similarly show that direct manipulation and constrained visual interaction can lower barriers and help novices build intuition about ML concepts \cite{carney2020teachable,smilkov2017playground}. These systems, however, typically foreground implementation and interaction style; they do not necessarily provide explicit guidance about the methodological adequacy of an experimental design.

A different strand of work addresses complexity through automation. AutoML systems reduce manual effort in model and hyperparameter selection by searching over candidate pipelines and configurations \cite{hutter2019automl,feurer2015autosklearn}. Yet research on human-centred AutoML also shows that automation introduces new interaction challenges around transparency, controllability, and trust. For example, ATMSeer demonstrates that users benefit from interfaces that allow them to inspect and steer the AutoML process, rather than merely receiving opaque results \cite{wangATMSeerIncreasingTransparency2019}. More broadly, guidelines for human--AI interaction emphasise the importance of exposing system capabilities and limitations, supporting user control, and calibrating expectations \cite{amershi_guidelines_2019,shneidermanBridgingGapEthics2020}. Recent copilot research points in the same direction: in feature-rich software, guided assistance can outperform fully automated assistance on perceived control and learnability, especially in exploratory tasks \cite{khurana2025copilots}. Taken together, this literature suggests that beginner-facing ML tools should not only reduce effort; they should also make reasoning and control visible.

This point is particularly important for methodological learning. Work on explanatory debugging argues that interactive intelligent systems can help users personalise and refine models when they expose reasons, support inspection, and invite corrective intervention \cite{kulesza_principles_2015}. Similarly, onboarding studies with medical practitioners show that domain experts want more than case-level predictions or explanations: they also want to understand a system's objectives, known strengths and weaknesses, and how it fits with their own standards of judgement \cite{cai_hello_2019}. Related work in HCI and IUI further suggests that interactive ML can help non-ML specialists connect model capabilities to application goals, but that such support needs to be scaffolded deliberately rather than assumed to emerge from access to tools alone \cite{fengAddressingUXPractitioners2023}. This literature motivates a distinction that, to our knowledge, has been underexplored empirically in beginner ML support: the difference between helping users produce a more plausible \emph{workflow structure} and helping them make better \emph{local methodological decisions} within that workflow.

\begin{itemize}
    \item[RQ1] What is the influence of tool-based assistance (structured vs unstructured) on the ML pipeline creators' conceptual understanding of ML pipelines?
    \item[RQ2] What is the influence of interactive programming environments (text-based vs visual) on the ML pipeline creators' task load, UX, and acceptance? 
\end{itemize}

In this paper, we study two complementary forms of support for beginner ML experimentation. The first is \textit{mlguide}, a structured and evidence-backed guidance component aimed at supporting conceptual workflow design. The second is \textit{AI Sandbox}, a visual no-code environment aimed at lowering implementation barriers by allowing users to compose workflows graphically. We evaluate these systems in two within-subject studies with predominantly technical beginners. In Study~1, participants first worked without structured guidance and then with \textit{mlguide}; we therefore treat the results as evidence about how workflow quality changes \emph{after} structured guidance, while acknowledging that the fixed order limits causal interpretation. In Study~2, we compare the visual environment with a notebook-based baseline, Google Colab, on user experience, workload, and acceptance.

Our central finding is that different forms of support appear to scaffold different layers of ML competence. Structured guidance improved participants' high-level workflow structure, but it did not measurably improve low-level methodological reasoning. The visual no-code environment, in turn, improved hedonic user experience relative to the notebook baseline, while workload and acceptance remained broadly comparable. We interpret these findings cautiously given the small sample, the participants' predominantly technical backgrounds, and measurement limitations. Nevertheless, the pattern is theoretically informative: it suggests that workflow structuring, methodological reasoning, implementation fluency, and user experience should be treated as related but distinct design outcomes for mixed-initiative ML tools.


We contribute the following:
\begin{enumerate}
    \item A \textbf{reusable task for ML workflow evaluation}, containing \textit{instructions} (see \autoref{app:briefing}), dataset modifications (see \autoref{app:dataset}), and an evaluating \textit{Quiz} (\autoref{app:task1.1} \autoref{app:task1.2}). This allows researchers for easy replication, inspiration, and extension of a way to test the understanding (high and low level) of ML work with near-to-real-world data with related issues that cannot be easily solved by AI agents.
    \item Empirical \textbf{insights into the effectiveness of tool-based structured guidance for ML workflow design} for non-experts' understanding (high/low level) through a controlled within-subjects experiment (see \autoref{sec:study_1}).
    \item Empirical \textbf{insights into implementation paradigms of ML workflow design tools}, i.e., task load and UX of text-based and visual programming environments for non-experts through a controlled within-subjects experiment (see \autoref{sec:study_2}).
\end{enumerate}
\section{Background \& Related Work}

Designing machine-learning (ML) experiments is not only a technical activity, but also an interaction problem. Users must decide which workflow steps are appropriate for a task, how these steps should be ordered, which methodological decisions are valid, and how the resulting workflow can be implemented in a concrete environment. Prior work has addressed parts of this problem through notebooks, visual workflow editors, AutoML systems, intelligent assistants, and educational interfaces. In this section, we use this literature to motivate a central distinction for our work: beginner-facing ML tools may support \emph{workflow structuring}, \emph{methodological reasoning}, \emph{implementation fluency}, and \emph{reflection} to different degrees. This distinction allows us to position structured guidance and visual no-code programming as complementary, but not interchangeable, forms of support.

\subsection{ML Workflow Design as an Interactive Intelligent Systems Problem}

ML development is typically iterative: practitioners inspect data, formulate modelling assumptions, prepare features, train models, evaluate alternatives, interpret results, and revise earlier decisions. Failures in this process often arise not only from model choice, but from weaknesses in the surrounding experimental workflow. Recent recommendations for ML-based science emphasise recurring problems around data leakage, inadequate validation, insufficient reporting, and limited reproducibility, and argue for stronger methodological standards across the full research and development process \cite{2024_Kapoor_REFORMSConsensusbasedRecommendations}. For beginner and occasional ML practitioners, this creates a dual challenge: they need to produce workflows that are executable, but also methodologically defensible.

In HCI, mental models describe the internal representations users form about how a system works and how actions lead to outcomes \cite{norman1983some,johnsonlaird1983mental}. When these models diverge from the actual system behaviour or from the structure of the task, users may struggle to anticipate consequences, choose appropriate actions, or recover from errors \cite{norman2013design,kieras1985mental}. In ML workflow design, we distinguish between two interrelated mental models. A \emph{conceptual model} concerns what a sound ML workflow should contain and why: for example, how data preparation, splitting, training, validation, comparison, and interpretation relate to each other. An \emph{implementation model} concerns how these steps are realised in a particular tool, programming environment, or visual workflow editor.

This distinction is consistent with prior work on interactive machine learning (IML), where interface design is central to how users inspect, steer, and understand learning systems. Dudley and Kristensson review user-interface design for IML and show that interactive ML systems vary in how they expose model behaviour, support user feedback, and make learning processes intelligible \cite{2018_Dudley_ReviewUserInterface}. More broadly, human--AI interaction guidelines emphasise that intelligent systems should make their capabilities and limitations clear, support user control, and help users recover from errors or unexpected behaviour \cite{amershi_guidelines_2019,shneidermanBridgingGapEthics2020}. For ML workflow tools, this implies that support should not only lower the effort of building pipelines, but also help users understand the methodological consequences of their choices.

\subsection{Computational Notebooks: Flexible Workspaces with Structural Burdens}

Computational notebooks are among the most common environments for data science and ML work. Jupyter popularised a document format that combines executable code, narrative text, equations, and visual outputs, thereby supporting both computation and explanation in a single workspace \cite{kluyver2016jupyter,jupyterdocs2026}. Google Colab extends the notebook paradigm as a hosted environment that requires little setup and provides access to computational resources useful for ML, such as GPUs and TPUs \cite{colabfaq2026,colabdev2026}. These properties make notebooks attractive for experimentation, teaching, and sharing.

However, HCI research also documents that notebooks impose substantial structuring work on users. Rule et al. show that notebook authors often balance exploratory analysis with the need to communicate and explain results, and that notebook documents can become difficult to navigate as analyses evolve \cite{rule2018notebooks}. Kery et al. describe how data scientists manage exploratory variants and how attempts to preserve the history of analysis can conflict with readability and presentation goals \cite{kery2018story}. Chattopadhyay et al. identify pain points around setup, hidden state, navigation, versioning, and reproducibility, showing that notebooks can make iterative work convenient while leaving important workflow dependencies implicit \cite{chattopadhyay_whats_2020}. Pimentel et al. further show that many notebooks in public repositories suffer from reproducibility problems, motivating stronger support for dependencies, execution order, and computational provenance \cite{pimentel2019repro}.

For our work, notebooks provide an important baseline: they offer high flexibility and are familiar to technically trained users, but they do not by themselves scaffold the conceptual structure of an ML experiment. Users must know which steps to include, where methodological pitfalls may occur, and how to translate those choices into correct code. This makes notebooks well suited for users with stronger implementation models, but potentially demanding for beginners whose conceptual models of ML workflows are still incomplete.

\subsection{Visual Workflow Systems: Externalising Pipeline Structure}

Visual workflow systems take a different approach by representing data-analysis and ML processes as explicit graphs of connected operators. KNIME is an early and influential example of a modular environment for visual assembly and execution of data-processing pipelines \cite{berthold2008knime,knimeDocs2026}. Orange similarly supports component-based data analysis through visual programming, while also offering scripting interfaces for more advanced use \cite{demsar2013orange,orangeDocs2026}. RapidMiner and its predecessor YALE also foreground reusable operators and visual process construction for data mining and predictive modelling \cite{mierswa2006yale,rapidminerDocs2026}.

The central advantage of such systems is representational. Pipeline stages, data transformations, and dependencies become visible as first-class objects. Compared with free-form code, visual graphs can make workflow order, composition, and intermediate artefacts easier to inspect. This can lower implementation barriers, especially when common preprocessing, modelling, and evaluation steps are encapsulated in reusable nodes or widgets. Visual programming is therefore not merely a different input modality; it changes which aspects of the workflow are externalised and available for reasoning.

At the same time, visual representation alone does not guarantee methodological adequacy. A visual workflow can make the order of operations visible, but it may still allow users to choose unsuitable validation strategies, inappropriate metrics, or misleading interpretation procedures. In other words, visual workflow systems can scaffold an implementation model and make workflow structure more explicit, but they do not necessarily teach why a particular workflow is methodologically appropriate. This is central to our paper: improving the visibility and manipulability of workflow structure may improve user experience and high-level workflow composition, while leaving low-level methodological reasoning largely unchanged.

\subsection{Automation, AutoML, and the Tension Between Efficiency and Control}

A third line of work reduces the burden of ML experimentation through automation. AutoML systems automate parts of algorithm selection, preprocessing, pipeline search, and hyperparameter optimisation \cite{hutter2019automl}. auto-sklearn combines scikit-learn with automated model selection, hyperparameter optimisation, meta-learning, and ensemble construction \cite{feurer2015autosklearn,autosklearnDocs2026}. TPOT frames pipeline construction as a genetic programming problem, automatically searching over preprocessing and modelling combinations to optimise predictive performance \cite{olson2016tpot,tpotRepo2026}. AutoML-Zero explores a more radical form of automation by searching for learning algorithms from low-level mathematical operations \cite{real2020automlzero,automlzeroRepo2026}.

These systems can reduce implementation effort and improve performance search, but they also shift the interaction problem. If users receive an automatically generated pipeline, they still need to understand what has been done, whether the result is valid for the task, and how much control they have over methodological constraints. Human-centred AutoML work therefore argues for transparency and controllability rather than automation alone. ATMSeer, for example, provides interactive visual support for inspecting and steering AutoML processes, showing that users benefit from seeing how candidate models are generated, evaluated, and compared \cite{wangATMSeerIncreasingTransparency2019}. This work is important for our positioning because it reframes AutoML as an interactive intelligent system: the goal is not only to find good models, but also to support human understanding and intervention.

Recent work on copilot-style assistance points to a similar design tension. Intelligent assistants can provide code suggestions, explanations, and task-specific recommendations, but the degree of automation matters. Khurana et al. compare more automated and more guided copilot paradigms in feature-rich software, showing that guided assistance can support perceived control and learnability, particularly in exploratory and creative tasks \cite{khurana_it_2025}. Conversational systems such as ChatGPT and coding assistants such as GitHub Copilot further illustrate how users increasingly seek flexible, on-demand help during programming and data-analysis work \cite{githubcopilotDocs2026,openaiChatGPT2026}. Notebook platforms also increasingly integrate AI assistance into code generation and transformation workflows \cite{colabdev2026}. Finally, systems such as InstructPipe explore the generation of node-graph pipelines from natural-language instructions, suggesting a convergence of visual workflow construction and language-based assistance \cite{zhou2025instructpipe}.

For beginner ML workflow design, this literature suggests that automation should not be treated as a single continuum from manual to automatic. Systems may automate search, recommend methods, generate code, visualise alternatives, or guide users through decisions. These forms of support have different implications for user control and learning. Our work positions structured guidance as a guidance-oriented alternative to full automation: rather than producing a finished pipeline for the user, it aims to structure the decision space and support more informed workflow composition.

\subsection{Learning, Onboarding, and Scaffolding for ML Beginners}

A separate but related body of work focuses on learning and onboarding. In the learning sciences, scaffolding describes temporary support that enables learners to perform tasks they could not yet complete independently \cite{wood1976tutoring}. Cognitive apprenticeship similarly emphasises modelling, coaching, articulation, reflection, and gradual transfer of responsibility \cite{collins1989cogapprentice}. Expertise research shows that novices and experts differ in how they categorise and represent problems, suggesting that beginners often need support in recognising structural regularities rather than only surface features \cite{chi1981expertnovice}. These ideas are relevant to ML workflow tools because beginners may not only lack commands or syntax; they may lack the conceptual structure needed to decide what the workflow should be.

Interactive educational ML tools provide constrained environments for building intuition. Teachable Machine allows users to train simple classification models through an approachable web interface without requiring specialised ML or programming knowledge \cite{carney2020teachable,teachablemachineSite2026}. TensorFlow Playground provides direct-manipulation visualisations of neural-network structure and hyperparameters, enabling users to experiment with model behaviour through immediate visual feedback \cite{smilkov2017playground,tfplaygroundSite2026}. These systems illustrate how simplified interaction and fast feedback can support intuition-building, but they are typically scoped to educational exploration rather than end-to-end methodological workflow design.

Work on explanatory debugging and domain-expert onboarding further shows that useful ML support must expose more than outputs. Kulesza et al. argue that explanatory debugging can help users personalise interactive ML systems when explanations support inspection and correction \cite{kulesza_principles_2015}. Cai et al. show that medical practitioners need onboarding support that explains not only model predictions, but also system goals, strengths, weaknesses, and appropriate use conditions \cite{cai_hello_2019}. Feng and McDonald similarly show that UX practitioners face difficulties when designing ML applications and benefit from interactive approaches that help connect ML capabilities to design goals \cite{fengAddressingUXPractitioners2023}. Together, these studies suggest that learning-oriented ML tools should support reflection and judgement, not only task completion.

This literature helps explain why structured guidance may improve high-level workflow composition without necessarily improving low-level methodological reasoning. A user may accept a recommended workflow step or assemble a more plausible pipeline without understanding the underlying statistical rationale. Deeper methodological learning likely requires additional mechanisms: explanations of trade-offs, examples and counterexamples, critique of user choices, opportunities to compare alternatives, and prompts that require users to justify why a method fits the task.

\subsection{Summary}

Prior work therefore supports several facets of beginner ML workflow design. Notebooks provide flexible and widely adopted implementation environments, but leave much workflow structuring implicit \cite{kluyver2016jupyter,rule2018notebooks,kery2018story,chattopadhyay_whats_2020}. Visual workflow tools externalise pipeline structure and lower implementation barriers, but do not necessarily ensure methodological soundness \cite{berthold2008knime,demsar2013orange,mierswa2006yale}. AutoML systems reduce search and configuration effort, but raise interaction challenges around transparency, control, and learning \cite{hutter2019automl,feurer2015autosklearn,wangATMSeerIncreasingTransparency2019}. Copilots and conversational assistants provide flexible help, but their benefits depend on whether they preserve user control and support understanding rather than merely producing outputs \cite{amershi_guidelines_2019,khurana_it_2025}. Educational and onboarding tools show the value of scaffolding, explanation, and reflection, but are often limited to simplified or domain-specific tasks \cite{carney2020teachable,smilkov2017playground,kulesza_principles_2015,cai_hello_2019}.

What remains underexplored is how different forms of interactive support affect different layers of beginner ML competence. In particular, prior work rarely separates improvements in high-level workflow structure from improvements in low-level methodological reasoning. This distinction is important because a workflow can look structurally plausible while still containing weak local choices, such as inappropriate validation, unsuitable metrics, or unjustified model comparisons. Our work addresses this gap by evaluating two complementary forms of support: structured, evidence-backed guidance for conceptual workflow design and a visual no-code environment for workflow implementation. Rather than asking only whether such tools make users faster or more satisfied, we ask which layer of ML workflow competence they scaffold.

\section{Design}

\subsection{Design Challenge and Rationale}
\label{sec:design-challenge}
\subsubsection{Current Practice Building in ML Workflow Design}
From the related work and our experience and understanding of ML workflow design, we characterize the design challenge by two main challenges of users knowing (1) what to do (\textbf{conceptual model}) and (2) how to do it (\textbf{implementation model}). For both, tools assist with building knowledge of how to approach (workflow guidance or method selection) and craft a solution (implementation).

\paragraph{Update/Assist Interaction Cycle}
Mental models also depend strongly on users’ expertise in a task: beginners often rely on simplified or incomplete representations of the workflow and its implementation, whereas experts tend to develop richer and more accurate models \cite{chi1981categorization}.
Designing ML workflows is a complex task with many parameters and users typically do not have a end-to-end solution from begin on because it is an iterative experimentation to find a good ML workflow solution.
To find missing pieces to the solution, users typically update missing pieces and information of their mental models by using tools that assist with updating their model with the right pieces.

Here, we distinguish two major user interactions, namely
(1), assistance for help updating the conceptual model (ML workflow guidance) and
(2) assistance for help updating the implementation model.
Most tool assist with either implementation or conceptualization, integrated approaches were sparse before the raise of LLMs. 
\begin{figure*}[ht!]
    \centering
    \includegraphics[width=\linewidth]{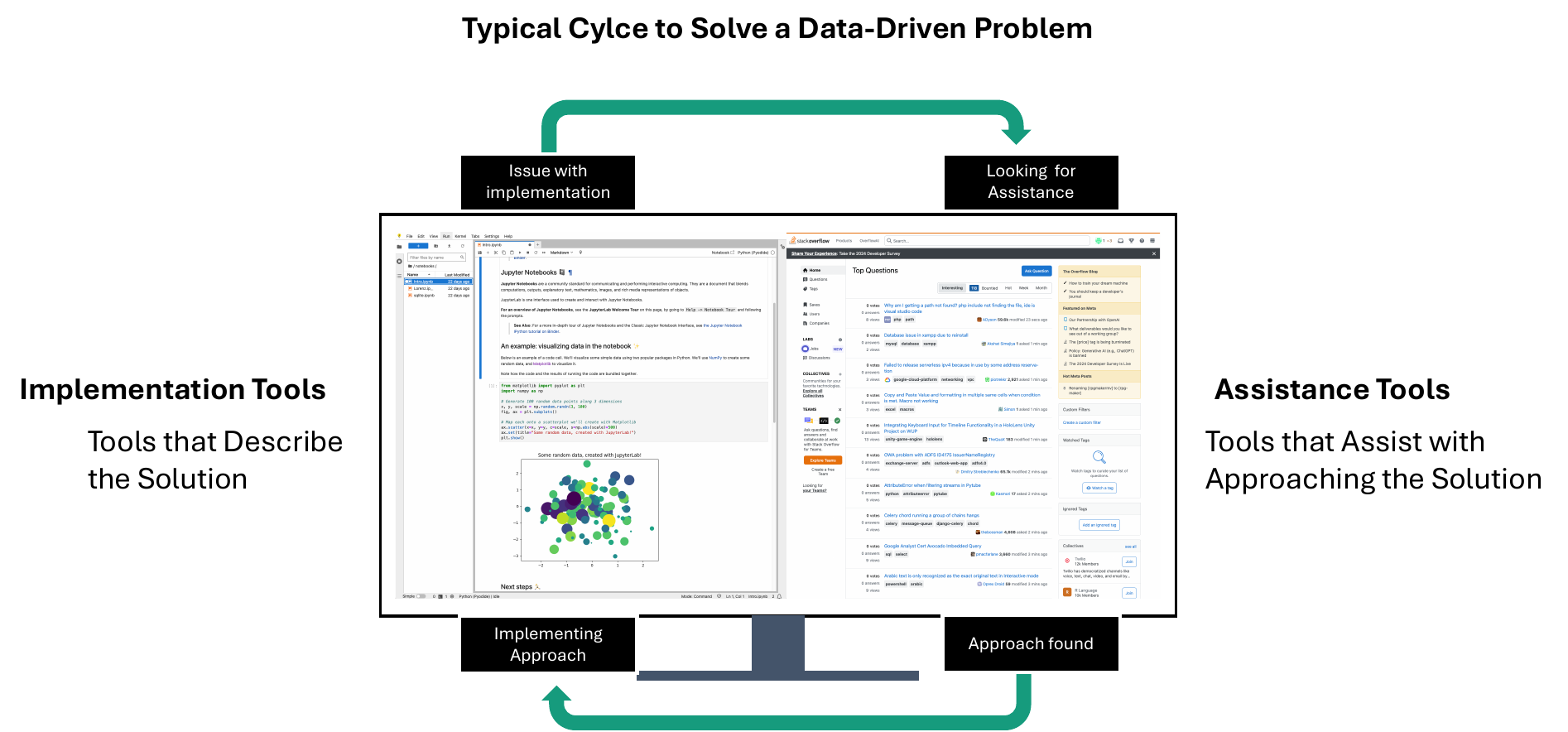}
    \caption{The update/assist interaction cycle in ML workflow design. A user encounters an implementation problem in a notebook (left), consults an external assistance tool such as Stack Overflow (right), and returns to the notebook to implement the selected approach.}
    \label{fig:problem}
\end{figure*}
Lowering barriers and reducing the missing pieces in their mental models is essential for ML workflow design tools.
Next, we briefly describe common barriers in ML workflow design.

\subsection{Challenges for the ML Workflow Design}

\paragraph{Methodological Challenges}
Designing machine learning (ML) workflows involves a range of methodological challenges related to data, modeling decisions, and evaluation procedures. Common pitfalls include data leakage, inappropriate validation strategies, class imbalance, insufficient sample sizes, or the misuse of evaluation metrics, all of which can lead to overly optimistic performance estimates or non-generalizable models. 
In addition, the iterative and exploratory nature of ML development makes it difficult to maintain methodological rigor and reproducibility throughout the workflow. 
Especially in critical domains such as healthcare, where ML model outcomes may have real-world consequences, careful ML workflow design and transparent documentation of modeling choices are essential to ensure validity and trustworthiness.

\begin{table*}[ht!]
    \centering
    \caption{Common methodological pitfalls in machine learning workflow design (non-exclusive!) – we synthesized a set of problems for beginners based on interviews by \citeauthor{2020_Kaur_InterpretingInterpretabilityUnderstanding}\cite{2020_Kaur_InterpretingInterpretabilityUnderstanding} and a paper by \citeauthor{2024_Lones_HowAvoidMachine}\cite{2024_Lones_HowAvoidMachine}.}
     
\begin{tabular}{p{3.8cm} p{12cm}}
\toprule
\textbf{Pitfall} & \textbf{Description} \\
\midrule
\noalign{\smallskip}

Insufficient Data Understanding and Quantity &
Lack of data exploration or insufficient data can lead to poor model generalization and unreliable results \cite{2024_Lones_HowAvoidMachine}. \\

\noalign{\medskip}

Duplicate Data &
Duplicate or inconsistently labeled data points can distort learned patterns and bias model predictions \cite{2020_Kaur_InterpretingInterpretabilityUnderstanding}. \\

\noalign{\medskip}

Redundant Features &
Including multiple representations of the same information can dilute feature importance and obscure model behavior \cite{2020_Kaur_InterpretingInterpretabilityUnderstanding}. \\

\noalign{\medskip}

Ad-hoc Feature Engineering &
Arbitrary transformations, such as binning continuous variables into categories, can lead to information loss and unstable model behavior \cite{2020_Kaur_InterpretingInterpretabilityUnderstanding}. \\

\noalign{\medskip}

Data Leakage and Improper Data Handling &
Information from test or validation data can unintentionally influence training or model selection, leading to overly optimistic performance estimates \cite{2024_Lones_HowAvoidMachine}. \\

\noalign{\medskip}

Missing Values &
Handling missing data (e.g., via mean imputation or encoding) can introduce bias, distort feature distributions, or lead to misleading model behavior \cite{2020_Kaur_InterpretingInterpretabilityUnderstanding}. \\

\noalign{\medskip}

Changes in Data Distribution &
Data distributions may shift over time or differ between training and deployment settings, leading to reduced model validity and generalization \cite{2020_Kaur_InterpretingInterpretabilityUnderstanding}. \\

\noalign{\medskip}

Inappropriate Model Choice and Spurious Learning &
Using unsuitable models or learning spurious correlations instead of meaningful patterns can lead to misleading predictions \cite{2024_Lones_HowAvoidMachine}. \\

\noalign{\medskip}

Debugging and Model Improvement Difficulties &
Identifying model errors or improvement opportunities based on limited observations or explanations is challenging, especially in complex workflows \cite{2020_Kaur_InterpretingInterpretabilityUnderstanding}. \\

\noalign{\medskip}

Improper Evaluation and Validation Practices &
Incorrect validation strategies, reuse of test data, or insufficient evaluation procedures can result in unreliable performance estimates \cite{2024_Lones_HowAvoidMachine}. \\

\noalign{\medskip}

Unsuitable Metrics and Model Comparison &
Choosing inappropriate metrics or omitting statistical comparisons can misrepresent model performance \cite{2024_Lones_HowAvoidMachine}. \\

\noalign{\medskip}

Misinterpretation and Over-Trust in Model Explanations &
Users may misunderstand or over-trust interpretability outputs, fail to correctly interpret visualizations, and draw incorrect conclusions from explanations \cite{2020_Kaur_InterpretingInterpretabilityUnderstanding}. \\

\noalign{\medskip}

Lack of Transparency and Reproducibility &
Insufficient reporting of modeling decisions and evaluation procedures limits reproducibility and trustworthiness \cite{2024_Lones_HowAvoidMachine}. \\

\noalign{\smallskip}
\bottomrule
\end{tabular}
    \label{tab:pitfalls}
\end{table*}
Designing ML workflows involves a range of methodological challenges related to data quality, modeling decisions, and evaluation procedures. 
Common pitfalls include data leakage, inappropriate validation strategies, class imbalance, insufficient sample sizes, or the misuse of evaluation metrics, all of which can lead to overly optimistic performance estimates or non-generalizable models \cite{kapoor_leakage_2023, sendak_presenting_2020}. 

These challenges primarily concern the \textit{conceptual model} of ML workflows (see \autoref{sec:design-challenge}). Users must understand how to construct appropriate ML pipelines and determine which methodological choices are suitable in a given context and why. 
However, the iterative and exploratory nature of ML development makes it difficult to maintain methodological rigor and reproducibility, particularly when decisions are made incrementally and without explicit documentation \cite{amershi_software_2019}. 
This is especially critical in high-stakes domains such as healthcare, where ML model outputs may directly influence clinical decisions, requiring careful workflow design and transparent reporting to ensure validity, reproducibility, and trustworthiness \cite{sendak_real_2020}. 

These challenges are further amplified by the gap between conceptual understanding and implementation. Even when users are aware of best practices, they may fail to correctly translate them into concrete workflow steps, such as properly separating data before feature engineering or selecting appropriate evaluation strategies \cite{kapoor_leakage_2023}. To structure this problem space, we compile a non-exhaustive set of common methodological pitfalls in ML workflow design in \autoref{tab:pitfalls}. These pitfalls span multiple layers, including data and workflow-related issues \cite{2020_Kaur_InterpretingInterpretabilityUnderstanding, 2024_Lones_HowAvoidMachine}, conceptual methodological challenges \cite{2024_Lones_HowAvoidMachine}, and human-centered interpretation and usage difficulties \cite{2020_Kaur_InterpretingInterpretabilityUnderstanding}.

\paragraph{Usability Challenges}

ML workflow design also poses significant usability challenges for users. 
These challenges largely arise from mismatches between users’ mental models of ML workflows and the system’s representation of the underlying processes (see previous \autoref{sec:design-challenge}). 
ML systems often require designing complex pipelines with numerous configuration options, or abstract concepts that may not necessarily align with users’ mental models of the ML process \cite{dove_monsters_2020, amershi_guidelines_2019}. 

Such mismatches particularly affect the \textit{implementation model} in a way that users may struggle to understand dependencies between workflow steps, anticipate the effects of parameter choices, or interpret intermediate and final results \cite{kulesza_principles_2015}. 
At the same time, limitations in the \textit{conceptual model} further nurture these issues, as users may lack the necessary abstractions to reason about concepts such as data leakage, validation strategies, or model evaluation \cite{domingos_few_2012, ng_machine_2018}. 

These challenges are especially pronounced for beginners, who often lack both well-formed conceptual representations and experience in mapping them to system operations. 
Prior work on end-user programming and learning sciences shows that novices struggle with abstraction, hidden dependencies, and transferring conceptual understanding into executable steps \cite{ko_six_2004, resnick_lifelong_2017}. 

Taken together, usability challenges in ML workflow design are not merely interface issues, but emerge from the interplay between conceptual understanding and implementation interaction. 
This reinforces our design problem: effective ML tools must support both the formation of appropriate conceptual models and their alignment with actionable implementation structures.

\subsection{Prototypes}

To address the needs of ML beginners and help with overcoming ML workflow design challenges, we developed \aisandbox which is available open-source\footnote{Repository will be linked upon publication.}.
\aisandbox is a no‑code environment that allows users to visually interact and design ML workflows, from data editing to running ML experiments. 
The environment provides an integrated end-to-end ML workflow toolbox with dedicated tools for data editing (\maratus), experiment design (\mlkitchen), and execution/evaluation (\mlflow). 
Additionally, it integrates workflow guidance (\mlguide) for rigorous method selection. 
Taken together, the tools lower the coding barrier (\maratus), provide transparent, evidence‑based guidance (\mlguide), make experiment design guided and error‑resistant (\mlkitchen), and ensure reproducibility and comparability by default (\mlflow).
While all tools work together in the \aisandbox environment, we separate between conceptual guidance (\mlguide) and implementation (\maratus, \mlkitchen, \mlflow) to better test our research questions.
Throughout this paper, when we refer to \aisandbox, we mean the required tools for ML workflow implementation tasks (\maratus, \mlkitchen, \mlflow), and \mlguide refers to the conceptual task with ML workflow guidance. 
We briefly describe the tools and how to use them for conceptualization/implementation tasks in the following.

\begin{figure*}[ht!]
    \centering
    \includegraphics[width=\linewidth]{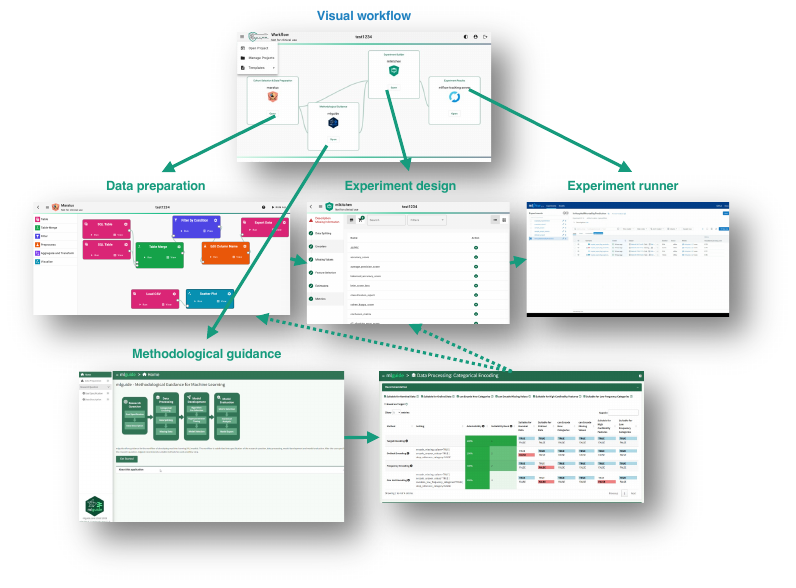}
    \caption{The visual ML experimentation workflow – The toolchain consists of a top-level tooling and project manager (\aisandbox, own development) and submodules (every part can also be done in Jupyter notebooks) for graphical low-barrier workflow design. The submodules include data preparation (\maratus, own development), experiment design (\mlkitchen, own development), and an experiment runner (\mlflow, open-source tool). Further, methodological guidance (\mlguide, own development) can be used for data preparation and experiment design to avoid typical pitfalls. Although integrated, in this paper, we study the guidance and implementation helpers in two subsequent experiments.}
    \label{fig:aisandbox}
\end{figure*}

\subsubsection{Implementation Task: \aisandbox}
The \aisandbox addresses users' implementation challenges.
The interface presents tools that users can drag onto a canvas and connect into a workflow.
There is a predefined flow: \maratus –(~modified~data~)$\rightarrow$ \mlkitchen –(~experiment~configuration~)$\rightarrow$ \mlflow.

\paragraph{\maratus}
\maratus allows to assemble data pipelines by providing drag\&drop and connect modules for loading, merging, filtering, transforming, aggregating, and visualizing tabular data. 
\maratus executes in chunks to handle larger tables efficiently and exports a single processed dataset plus metadata (feature types, missingness, simple stats) that can be used by \mlkitchen.
For beginners, the visual pipeline makes data preparation concrete and reduces error‑prone, code‑heavy steps. Through the saved metadata, users do not need to remember technical details, e.g., \mlkitchen later auto‑configures encoders/imputers.

\paragraph{\mlkitchen}
\mlkitchen allow to design an executable ML experiment without coding, optionally pre-filled via \mlguide's recommendations. 
It walks users step‑by‑step through the pipeline: splitting, optional imputation and encoding (pre‑seeded using \maratus metadata), optional feature selection, estimator selection with optional hyperparameter tuning, and metrics. 
For beginners, this reduces cognitive load: everything is visible, ordered, and reproducible, and guardrails (like splitting first) are built in to prevent leakage.
We provide sensible defaults and documentation to help users understanding parameters. 
\mlkitchen produces a ML experiment configuration \textit{recipe} as output which is then mapped to \mlflow configurations (ingest, split, transform, train).

\paragraph{\mlflow}
\mlflow executes and tracks the designed experiments. 
From \mlkitchen's recipe, \mlflow runs the experiment, and logs configs, code, metrics, and artifacts. 
Finally, users can check their models performance (over time).
Beginners benefit from automatic versioning and comparisons across runs, which encourages iterative learning.

\subsubsection{Conceptualization Task: \mlguide}
The \aisandbox addresses users' methodological challenges.
\mlguide provides evidence‑based, transparent methodological guidance.
Users describe their problem (e.g., classification vs. regression, class balance, dataset size, constraints), optionally prefilled from maratus metadata.
The guidance engine is built on an explicitly curated evidence base: rules extracted from the literature encode statements like ``Method M has property P under condition C with belief B.''
This evidence is presented to users via suitability rankings in \textit{recommendation tables}.
For beginners, \mlguide teaches best practices on the spot, e.g., avoiding data leakage, choosing stratified splits for imbalance, picking metrics aligned with the goal. 

In contrast to conversational assistants such as \textit{ChatGPT} or community-based sources such as \textit{StackExchange}, \mlguide does not primarily provide open-ended answers, but a structured, evidence-backed representation of design space. This positions \mlguide closer to guidance-oriented assistant paradigms than to automation-first systems. 
\section{User Studies}
To answer how to assist users with (1) conceptualization and (2) implementation of ML workflows, we conducted two subsequent within-subject experiments (N=18), each focusing on one of these aspects.

\subsection{General Procedure}
The survey starts with briefing and informed consent.
For the first aspect, conceptualization, we provide participants with a prepared dataset and let them submit an ML workflow concept while allowing any kind of help such as LLM assistance via \textit{ChatGPT}. We assess the workflow quality by comparing their submission with a predefined workflow and asking for a rationale for individual workflow steps. 
Next, participants use \mlguide for structured design guidance and adapt their submission. We assess the effectiveness of the structured guidance through the workflow concept changes and user experience. 

\begin{figure*}[ht!]
    \centering
    \includegraphics[width=.75\linewidth]{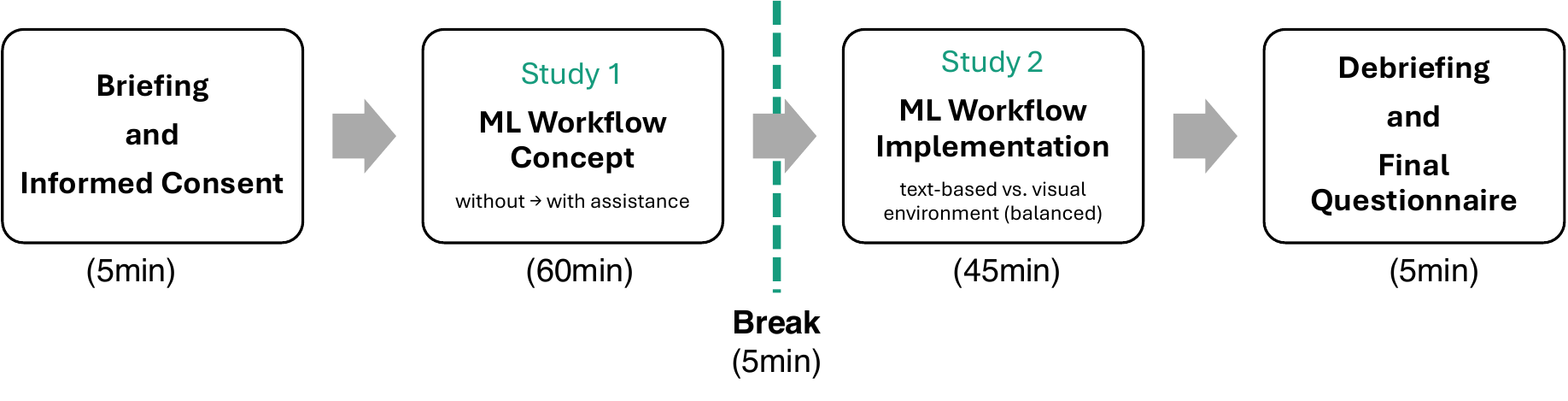}
    \caption{Overview of the \textbf{general procedure of the survey}. After briefing and initial consent, participants worked for an hour on the conceptualization of an ML workflow design (around 30min without and 30min with structured assistance through \mlguide). Then, they tested visual and text-based approaches to the ML workflow implementation in a randomized (counter-balanced) order with the tools \colab (text-based) and \aisandbox (visual). Finally, participants finished with a socio-demographics questionnaire, could leave general remarks, and are debriefed.}
    \label{fig:overall_procedure}
\end{figure*}

After a short break, we continued investigating the second aspect, implementation. In this experiment, participants rate their task load, acceptance, and UX for an implementation task using either a visual or text-based interactive programming environment. Finally, participants fill out a questionnaire that assesses socio-demographics and general remarks, before they are debriefed. \autoref{fig:overall_procedure} shows an overview of the process.

In the following, we describe the studies' setup and sample, before we present their method and results in separate sections (cf. \autoref{sec:study_1} and \autoref{sec:study_2}) for better readability. Finally, we discuss both jointly (cf. \autoref{sec:discussion}).

Before the studies, we sent out an Email that informed participants about the studies' purpose, procedures, compensation (25€ for $\sim$2h), and their right to withdraw at any time without penalty – according to best practices of the GDPR~\cite{eu_gdpr} and Declaration of Helsinki~\cite{WMA2013}. 
The informed consent statement emphasized the confidentiality of their data and the use and storage for research purposes only.

\subsection{Setup}
Since all tested tools are web applications, we offer remote and on-site participation. 
For deployment of the tested tools \mlguide and \aisandbox, we use a company-internal, cloud-based server environment that provides access to predefined user accounts via \textit{keycloak} \cite{wildfly_keycloack_2024}.
Regarding hardware, on-site, we provide 13''-screen laptops on single desks in a seminar room, while remote participants use their own devices and connect via Zoom\footnote{\url{http://zoom.us}}. 
As a Browser, all participants are instructed to use \textit{Google Chrome}\footnote{\url{https://www.google.com/chrome}}.
The study itself and the materials are hosted on \textit{SoSciSurvey}\footnote{\url{https://www.soscisurvey.de}}.

\subsection{Analysis}
\label{sec:analysis}
To identify a sufficient number of study participants, we conduct apriori power analysis using \textit{G*Power v3.1.9.6} \cite{faul_gpower_2007} to estimate the sample size, assuming at least a small effect ($dz=0.3$) size of our design interventions according to Cohen \cite{cohen_statistical_2013}. With a significance level of $\alpha = .05$ and a power of .80, we determine that the minimum required sample size for this effect size is $N = 16$ for a repeated measures t-test (or similar linear model) that sufficiently tests our studies' hypotheses. 
We conduct the analysis using \textit{R-script} \cite{rcoreteam_language_2021} using interactive notebooks (see also supplementary materials).
We create descriptive plots with ggstatsplot \cite{ggstatsolot2021}.
We use a linear mixed-effects model (short: LMEM, implemenation of packages lme4 \cite{lme4-jss} and lmerTest \cite{lmerTest-jss}) to test the influence of condition \textit{condition} (Experiment 1: time pre/post \mlguide use; Experiment 2:  \colab/\aisandbox used) and participants’ \textit{experience level} (abbreviated with \textit{xp}, see \autoref{sec:participants}) on the dependent variable (\textit{score}). The model includes random intercepts for participants ($1|\text{ID}$): 
\begin{equation}
    score \sim condition * xp\_score + (1|ID)
\end{equation}

One can find the complete data analysis script in the supplementary materials.

\subsection{Participants}
\label{sec:participants}
We recruited 18 participants via University, company, and personal networks, of whom identified twelve as male, three as female, and one as non-binary, two do not disclose. Nine conducted the studies on-site and another nine remote participated remote.
The average age is of participants is 25.19±3 years. Most of the participants are students and a few are professionals in a field where ML workflow design is of high relevance such as MSc. in \textit{Artificial Intelligence and Intelligent Systems} or {Digital Media Software Developer}. Three fall not into these programming/ML-near categories: one did not disclose, one is unspecific (\textit{student}), and one somewhat far from programming/ML (\textit{International Relations}).
\autoref{app:participants} shows per-participant data.
We compensated participants for the total two-hour procedure of the two experiments with 25€. Prerequisites for the studies are basic knowledge of ML and Python programming.

\subsubsection{Experience}
To account for prior ML and programming experience, we asked participants: ``How many years of experience do you have in programming and machine learning in academic and/or industrial settings?'' Further, we ask for a self-assessment of competence on a six-point bi-polar scale using four items for ML/programming conceptual and implementation experience level:
\begin{itemize}
    \item Conceptual experience level: (1) ``I just started learning about [programming/ML] concepts.'' $\leftrightarrow$ (6) ``I know most of the important [programming/ML] concepts.''
    \item Implementation experience level: (1) ``I do [programming/ML] seldom.'' $\leftrightarrow$ (6) ``[Programming/ML] is my daily business.''
\end{itemize}

The four items show strong correlations between conceptual and implementation experience ($r$=.80) and moderate correlations between ML and programming experience ($r$=.40).  Therefore, we aggregated all items into a single continuous experience score ($xp\_subjective\_overall$; Cronbach's $\alpha$=.80), combining subjective competence and averaged years of practice to approximate overall technical expertise ($years\_ML$, $years\_programming$):
\begin{equation}
\label{eq:xp}
\begin{aligned}
\mathrm{xp\_score}
  &= \mathrm{xp\_subjective\_overall\_pct} \\
  &\quad {}*
    \frac{\mathrm{years\_ML} + \mathrm{years\_programming}}{2}
\end{aligned}
\end{equation}

\begin{table}
\centering
\caption{Participants' practical/conceptual programming and ML experience. We compute a general technical experience score.}
\begin{tabular}{l c}
\toprule
\textbf{Experience Metric} & \textit{Mean$\pm$SD} \\
\midrule
Overall ML/programming (\%)               & $.61\pm.18$ \\
\quad Conceptual ML/programming (\%)    & $.65 \pm .17$ \\
\quad Implementation ML/programming (\%) & $.58\pm.21$ \\

Years of programming     & $3.78\pm1.96$ \\
Years of ML              & $1.50\pm1.25$ \\
\hline
\hline
\textbf{xp score} (cf. \autoref{eq:xp})            &  $1.76\pm1.10$\\
\bottomrule
\end{tabular}
\label{tab:xp}
\end{table}

We included this overall score as an experience moderator in the analysis (see previous \autoref{sec:analysis}).
Thus, moderation effects should be interpreted as reflecting general technical expertise across conceptual and implementation dimensions.
Next, we report on the individual experiments.
\section{Study 1 – Assisting ML Design with Structured Guidance} 
\label{sec:study_1}

\begin{figure*}[ht!]
    \centering
    \includegraphics[width=0.99\linewidth]{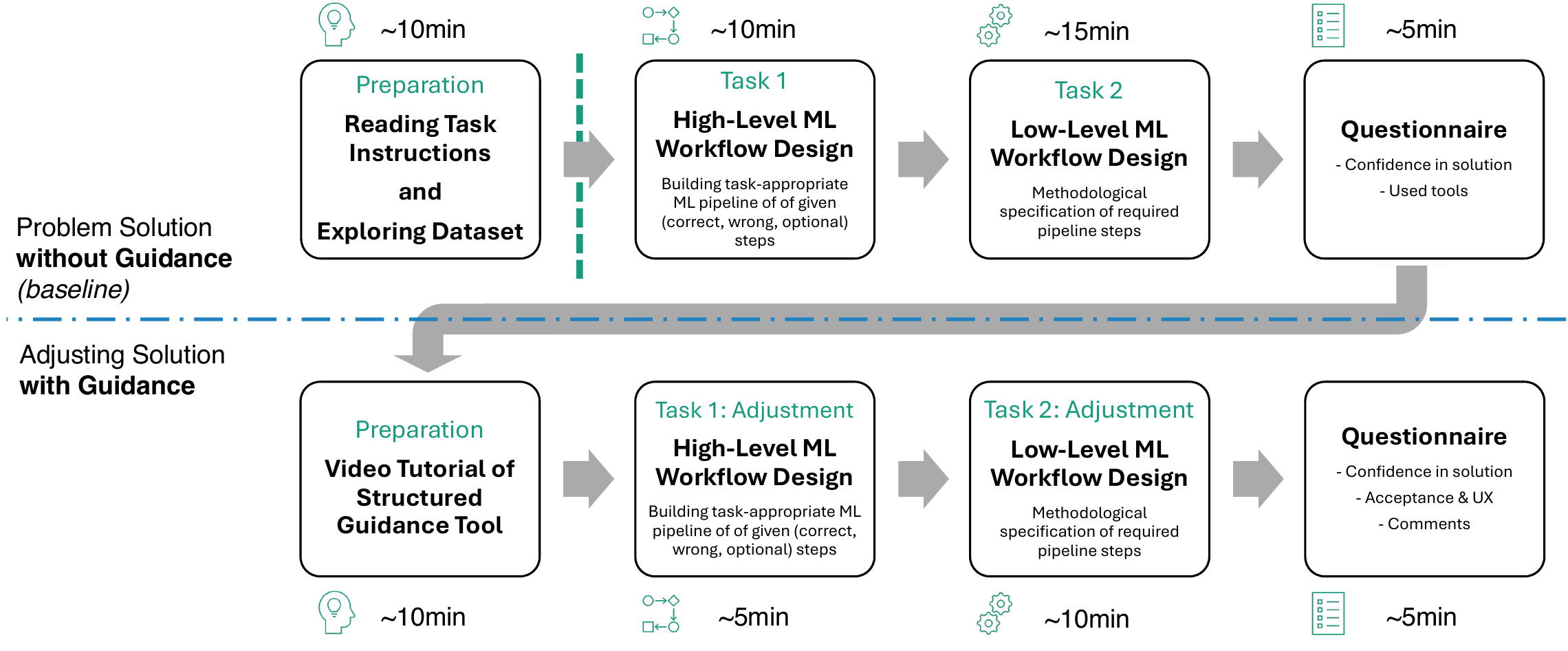}
    \caption{Overview over the procedure of Study 1.}
    \label{fig:procedure-1}
\end{figure*}

To answer how to assist users with the conceptualizations quality and rigor of ML workflows, we employed a two-phase within-subject experiment.
First, participants should solve a medical ML workflow design task (free to use any tool) by inspecting a given (manipulated) dataset and generating a suitable pipeline and choosing the right methods.
Then, they could use the assistance tool \mlguide and correct their solutions. 
We compared tasks and edits in terms of outcome (high-level workflow design) and methodological understanding improvements (low-level workflow design) as well as and acceptability and UX of \mlguide.
To replicate this study or to use it for similar ML workflow design tasks in the future, especially when investigating ML beginners, we provide an in-depth description and supplementary materials for dataset, task, and evaluation.
We present the full method in the following (see also \autoref{fig:procedure-1}).

\subsection{Method}
\label{sec:study_1_method}

\subsubsection{Design}
We used a two-phase mixed factorial study design with the independent variable \textit{Workflow Guidance Modality} (Unstructured Guidance vs. Structured Guidance).
The order was fixed to \textit{Unstructured Guidance}$\rightarrow$\textit{Structured Guidance}. So that we could measure the task baseline (using state-of-the-art-tools) first, before the intervention took place and avoid bias through the structured approach.
Both conditions required to solve a ML workflow problem (see \autoref{sec:materials}). 
The design allows for direct within-participant comparisons of design decisions and outcomes.
First, for the condition Unstructured Guidance, participants could use was an unassisted ML workflow task solving, as participants would do it normally, i.e., all kinds of aid were allowed including tools like LLM-chats, or community-plattforms like stackexchange\footnote{\url{https://stats.stackexchange.com}}.
Second, for the condition Structured Guidance, participants used the tool \mlguide in addition to their normal workflow and could have a second look and correct their initial design. \mlguide was not explicitly tailored to solve the given task rather this test was an initial validation of the tool itself. \mlguide can aid with a variety of tasks, e.g., in this setting it could not point users to the right answer yet provided them with ranked recommendations for algorithms for different workflow phases or provided checkboxes for data-related problems based on the initial problem definition which is up to the participants.

The dependent variables were 
(1) high-level workflow design,
(2) low-level workflow design,
and (3) subjective experiences during the task (UX and acceptance).
Regarding the first point, high-level workflow design, 
we measured the workflow quality via completeness and correctness of the designed ML pipeline. 
Regarding the second point, low-level workflow design, 
we measured methodological rigor (e.g., appropriate data splitting, metric choice, explainability of the chosen model).
Regarding the third point, subjective experiences, 
we measured confidence in first and second solution, UX and acceptance of \mlguide. 
We detail out the workflow assessment in the corresponding subsections (\autoref{sec:study_1_method_task}, \autoref{sec:study_1_method_measurement}).

\subsubsection{Materials}
\label{sec:materials}

The study scenario simulated a realistic hospital data-science task (see \autoref{sec:study_1_method_task}) and evaluation (see \autoref{sec:study_1_method_measurement})in which participants were asked to design an ML workflow for stroke-risk prediction under clinical constraints (e.g., higher cost of false negatives, demand for calibrated confidence estimates, and explainability requirements). 
To include typical data science pitfalls, we prepared a stroke prediction dataset (see \autoref{sec:study_1_method_dataset}). 

\paragraph{Dataset}
\label{sec:study_1_method_dataset}
The study dataset is based on the public Stroke Prediction Dataset from Kaggle \cite{dataset-orginal}.
The original dataset contains 12 columns, including the record identifier and the binary stroke outcome. The script adds three columns, modifies BMI missingness, and appends 100 sampled rows as exact duplicates.
\begin{table*}[ht!]
    \centering
    \caption{Stroke dataset fields and study modifications. The final column names the corresponding challenges in \autoref{tab:pitfalls}. The script also reordered columns for presentation.}
    \label{tab:dataset-features}
    \begin{tabular}{@{}p{3cm}p{4cm}p{9.5cm}@{}}
        \toprule
        Field & Description & Modification for the study \\
        \midrule
        \texttt{id} & Record identifier & \emph{Duplicate data.} Appended 100 exact copies of sampled rows, retaining their IDs. \\
        \noalign{\smallskip}
        \texttt{gender} & Reported gender & -- \\
        \noalign{\smallskip}
        \texttt{age} & Age in years & -- \\
        \noalign{\smallskip}
        \texttt{age\_category} & Age band & \emph{Ad-hoc feature engineering; redundant features.} Binned age into ten left-closed intervals and retained the original age. \\
        \noalign{\smallskip}
        \texttt{hypertension} & Binary health indicator & -- \\
        \noalign{\smallskip}
        \texttt{heart\_disease} & Binary health indicator & -- \\
        \noalign{\smallskip}
        \texttt{ever\_married} & Marriage indicator & -- \\
        \noalign{\smallskip}
        \texttt{work\_type} & Employment category & -- \\
        \noalign{\smallskip}
        \texttt{work\_type\_num} & Integer employment code & \emph{Redundant features.} Integer-encoded the five employment categories and retained the original field. \\
        \noalign{\smallskip}
        \texttt{Residence\_type} & Urban or rural residence & -- \\
        \noalign{\smallskip}
        \texttt{avg\_glucose\_level} & Average blood glucose & -- \\
        \noalign{\smallskip}
        \texttt{bmi} & Body mass index & \emph{Missing values; data leakage.} Set BMI to missing for 200 sampled stroke-positive records. Filled missing BMI for stroke-negative records with draws from an empirical BMI distribution. The resulting missingness reveals the outcome. \\
        \noalign{\smallskip}
        \texttt{smoking\_status} & Reported smoking status & -- \\
        \noalign{\smallskip}
        \texttt{stroke} & Binary prediction target & -- \\
        \noalign{\smallskip}
        \shortstack[l]{\texttt{survival\_3\_months\_}\\\texttt{post\_stroke}} & Synthetic three-month survival & \emph{Data leakage and improper data handling.} Added for stroke-positive records: 0 for 30 randomly selected cases and 1 for the rest. Left blank for stroke-negative records. This field leaks the target. \\
        \bottomrule
    \end{tabular}
\end{table*}

We modified the dataset to introduce common methodological pitfalls (\autoref{tab:pitfalls}) and make the task less directly solvable from familiarity with the original dataset. \autoref{tab:dataset-features} lists the changes, \autoref{app:dataset} shows an excerpt, and the supplementary materials contain the full modified dataset and the script.
The script directly introduces examples of five challenges from \autoref{tab:pitfalls}: duplicate data, redundant features, ad-hoc feature engineering, data leakage and improper data handling, and missing values. It also creates opportunities for spurious learning through outcome-linked fields and missingness. Whether a model learns those shortcuts depends on the subsequent workflow, and the script itself does not select or train a model.

The script does not explicitly manipulate data understanding or quantity (the duplicate rows add no independent observations), simulate distribution shift between training and deployment, or introduce challenges involving debugging and model improvement, evaluation and validation, metric selection and model comparison, interpretation and over-trust, or reporting and reproducibility. Some of these challenges are addressed separately in the study tasks.

We did not provide the whole dataset to participants but an excerpt and column-wise summaries. 
This data excerpt and summary served as a base for the task that we gave to the participants.

\paragraph{Task}
\label{sec:study_1_method_task}
We designed a task that resembles a typical real-world data science task in healthcare.
Participants should imagine working as data scientist in a hospital's stroke unit.
We ask them to create a prediction model and design a ML workflow for predicting stroke events from the given dataset (see \autoref{app:dataset}). 
The task required to 
(1) design an \textit{ML workflow}, and
(2) select \textit{methods} for some selected workflow steps.

We provided an overview of how to approach the task.
Participants should
(1) prepare the data for training and evaluation, 
(2) select and train a suitable ML model,
(3) evaluate model performance (constrains: (a) false negatives are more severe than false positives, and (b) model outputs a reliable prediction confidence),
(4) compare their model to a baseline model, and 
(5) choose an approach that interpretable and understandable for physicians.
The concrete task instruction are in \autoref{app:briefing}.

\paragraph{Measurement}
\label{sec:study_1_method_measurement}

To measure the task performance (high-level/low-level workflow design), we created a quizzes with a scoring system and for subjective experiences (taskload, UX and acceptance), we used questionnaires. We briefly describe the three main measuring dimensions in the following.

First, the quiz for the high-level workflow design quiz is a sorting procedure.
Participants can drag and drop predefined workflow steps on up to 20 slots a canvas which represent the workflow execution order and test the participants' ability to create a end-to-end ML experiment.
The predefined workflow steps consisted of task appropriate steps, optional steps, and wrong steps (distractions) -- see \autoref{app:task1.1} for the complete list.
We measured the ML workflow quality via completeness and correctness criteria (see also \autoref{app:task1.1}).
For every workflow step participants scored points, ranging from +2 (required and correct step selected) to  -1 (distraction selected, step missing, ordering violations), so that we could calculate an overall high-level workflow design score for every participant.

Second, the low-level workflow design quiz requires six multiple/single choice answers for five required workflow steps.
Here, we measured methodological rigor and the in-depth understanding of ML workflow design of participants, i.e., what methods to use for the workflow steps and how well the methods fit to the task description.
For example, choosing and appropriate data splitting strategy or choosing models that ensure explainability for clinicians.
In specific, we picked 5 workflow steps for which we ask participants in-depth methodological questions:
(1) data preprocessing / data cleaning (2 questions), 
(2) data splitting, 
(3) metric selection, 
(4) algorithm preselection, and
(5) statistical analysis.
We measured the ML workflow rigor via completeness and correctness criteria (see \autoref{app:task1.2}).
For every question we scored points, depending on questions type (up to 6 points), so that we could calculate an overall low-level workflow design score for every participant.

Third, the subjective experiences includes confidence in initial/edited solution, UX and acceptance of \mlguide. 
To measure the confidence in their solution, we asked a single question with ``I am confident with my [latest] approach.'', and, to measure acceptability, we used a single item ''I would use \mlguide for my academic/industrial machine learning workflow designs.''
Both items used a 7-point Likert scale (endpoints: "fully disagree"/"fully agree").
``I would use mlguide for my academic/industrial machine learning workflow designs.''
We employ UX via UEQ-S \cite{schrepp_design_2017} (semantic differential scale from -3 to +3).

In addition, we also asked which kind of help participants had used for their initial solution: Google Search (or similar, e.g., Bing, DuckGogo, ...), StackExchange (includes StackOverflow and similar), ChatGPT (or similar LLM-based support), and free text. They could express the subjective impression of the contribution of the assistance ("very low impact"/"very high impact" endpoints on a 7-point Likert scale).

\subsection{Procedure}
\autoref{fig:procedure-1} shows a visual outline of the procedure.
The study consists of two subsequent phases: A baseline solution where participants could use any kind of help (Unstructured Guidance) and an intervention follow-up where they could adjust their solutions after being introduced to the tool \mlguide (Structured Guidance).
Before the baseline, participants got briefed about the procedure and their upcoming task by the experimenter, they explore the dataset and ask clarifying questions (10min).
Instructions were printed (on-site) or provided as pdf (online) so that participants could always access the dataset excerpt and task instructions.
After initial instructions, the time started and was tracked by the experimenter to keep the whole study task duration to about 1 hour.
We instructed participants that the timing is not fixed for specific tasks, but after half an hour (and five minutes earlier), the experimenter will announce that is time to finish the first part (skip the quizzes and go to the questionnaires) and move on to the second part. In the second part, the experimenter announced that is time to finish up five minutes before the end. 

In the first part, participants solved (1) high-level workflow design quiz (10min), (2) the low-level workflow design quiz (15min), (3) rated their confidence in the solution, and (4) listed their aids (5min).
In the second part, we participants them an tutorial video for the use of \mlguide (10min) and they could input their problem description into \mlguide and they were instructed to (1/2) correct/edit their initial answers in the quizzes (25min), before they assessed their subjective experience (5min). Subjective experience included (3) the latest confidence in the updated solution, (4) UX (UEQ and free text), and (5) acceptability of \mlguide.
After that, a participants had a break, before they did the second study.
Next, we will present the how different Workflow Guidance Modalities impacted the task outcome and experience of users.

\begin{figure*}[t]
    \centering
    \includegraphics[width=0.62\linewidth]{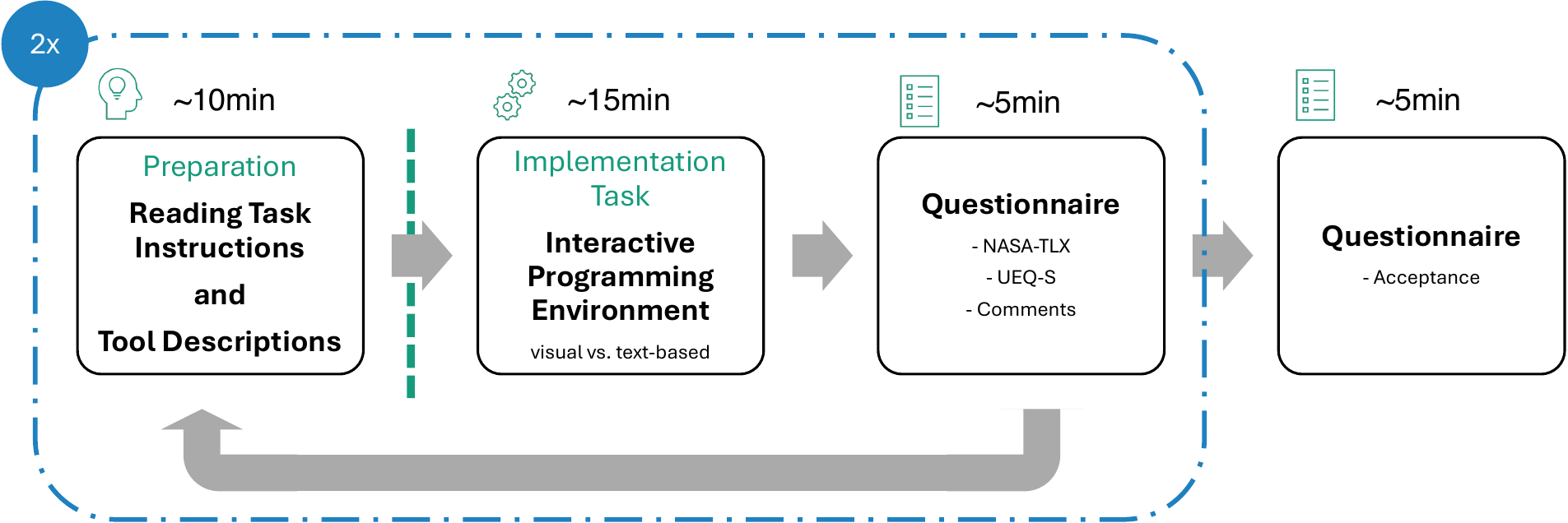}
    \caption{Procedure of Study 2.}
    \label{fig:procedure-2}
\end{figure*}

\subsection{Results}

\paragraph{High-Level Workflow Design}
After using \mlguide, pipelines are scored significantly better on overall structure ($\beta$=2.11, t(16)=3.93, p$\leq$.001, d=0.98, CI$_{95}$[0.42,1.54]), from (M=5.67$\pm$3.36) pre-guide to (M=7.78$\pm$4.40) post-guide. 
There is also a significant time x experience interaction, showing that this improvement is stronger for more experienced users ($\beta_{\text{time:xp}}$=1.53, t(16)=3.06, p=.007, d=0.76). 
To sum it up, the guide clearly enhanced pipeline design quality towards best practices, especially for more experienced users.

\begin{figure*}[ht!]
    \centering
    \begin{subfigure}[b]{0.41\textwidth}
        \includegraphics[width=\textwidth]{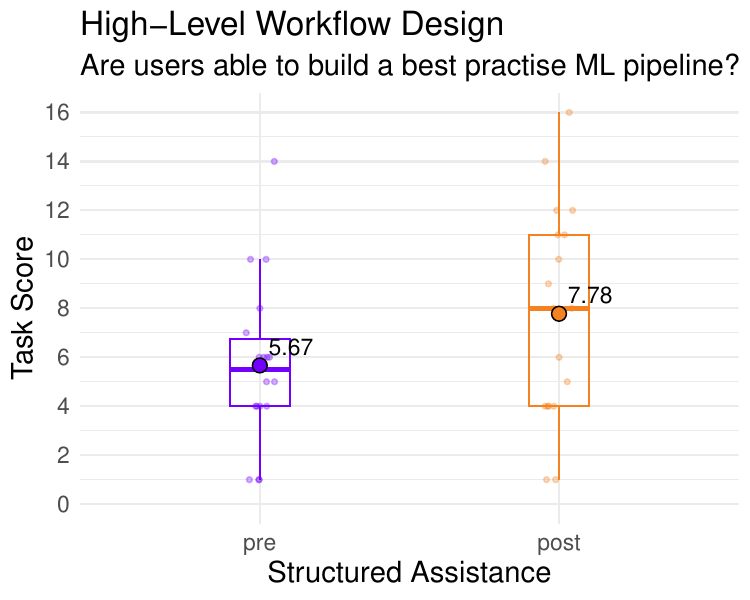}
        \caption{High-Level Workflow Design}
        \label{fig:performance}
    \end{subfigure}
    \begin{subfigure}[b]{0.58\textwidth}
        \includegraphics[width=\textwidth]{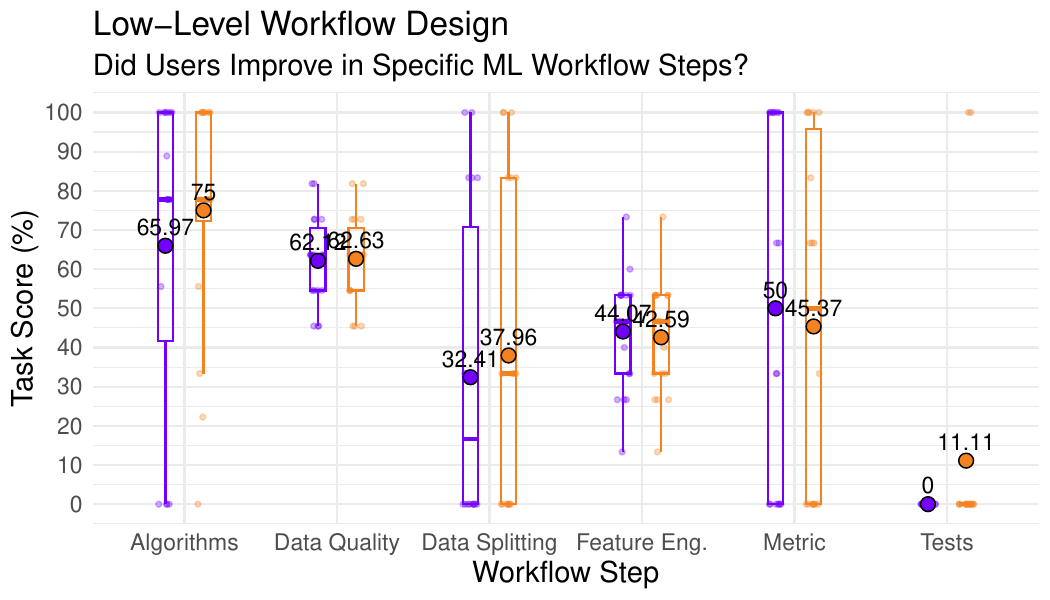}
        \caption{Low-Level Workflow Design}
        \label{fig:low-level-design}
    \end{subfigure}
    \begin{subfigure}[b]{0.36\textwidth}
        \includegraphics[width=\textwidth]{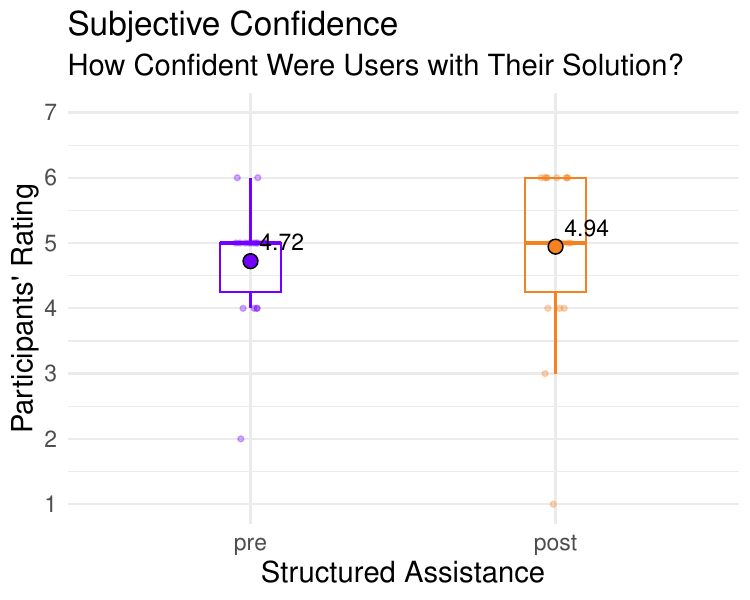}
        \caption{Confidence}
        \label{fig:confidence}
    \end{subfigure}
    \caption{Task-related measures in Study 1. (a) High-level workflow design scores before and after using \mlguide. (b) Low-level methodological quiz scores before and after using \mlguide. (c) Participants\textquotesingle{} confidence in their workflow solutions before and after using \mlguide.}
    \label{fig:study1-task-measures}
\end{figure*}

\paragraph{Low-Level Workflow Design}
After using \mlguide, we found no significant improvements in participants' quiz answers (selecting the right methods) to the six fundamental workflow steps (Data Quality, Feature Engineering, Data Splitting, Metric, Algorithms, Tests) of our use case. 
\mlguide use does not lead to better fine-grained workflow design decision, e.g., participants do not choose the right evaluation metrics more often. See also \autoref{fig:low-level-design}.

\paragraph{Confidence}
\autoref{fig:confidence} shows the confidence ratings. Confidence increases slightly by on average 0.23 points (4.72$\pm$0.90$\rightarrow$4.94$\pm$1.35) post \mlguide use, but the change is non-significant ($\beta$=0.22, t(16)=0.99, p=.34, d=0.24, CI$_{95}$[–0.24,0.70]).

\paragraph{User Experience \& Acceptance}
Regarding the UEQ-S, participants perceive \mlguide's pragmatic (M=0.78$\pm$0.98) and hedonic qualities (M=1.06$\pm$0.83) rather positive.
Further, participants found \mlguide acceptable (M=5.67$\pm$1.08), i.e., would use it for their ML workflow design in academia/practice.

\section{Study 2 – Comparing Interactive Programming Environments}
\label{sec:study_2}

To investigate the effect of the interactive programming environments modality (text-based/visual) on creators' task load, UX and acceptance, we conducted a crossover within-subjects experiment.
Participants should implement workflow steps using either \aisandbox or \colab.
In this experiment, we are interested in the experience of beginners regarding perceived load, UX, and acceptability.
We present the full method in the following (see also \autoref{fig:procedure-2}).

\subsection{Method}

\paragraph{Design}
We used a crossover mixed factorial study design with the independent variable \textit{Interactive Programming Environment} (text-based vs. visual).
The presentation order was counterbalanced.
In both conditions, participants should solve a simple task that required some engagement with the tools.
We designed tasks identically and included simple instructions to follow covering data loading, feature editing, training and evaluation (see \autoref{sec:study_2_method_task}). 
The baseline condition is the text-based environment represented through the tool \colab and the visual environment is represented through our developed tool \aisandbox.
We chose \colab as the text-based baseline because it provides a hosted Jupyter-style notebook for combining executable code, instructions, and outputs \cite{kluyver2016jupyter,colabfaq2026}. For context, 11.4\% of respondents to the IDE question in the 2026 Stack Overflow Developer Survey reported regularly using Jupyter Notebook or JupyterLab \cite{stackoverflow2026devide}.
The design allows for direct within-participant comparisons of experiences with the different interactive programming environments. 
The dependent variables were the subjective experiences during the task (taskload, UX and acceptance).
We detail out the experience assessment in the corresponding subsections (\autoref{sec:study_2_method_task}, \autoref{sec:study_2_method_measurement}).

\paragraph{Task}
\label{sec:study_2_method_task}
To get participants involved into the two programming environments, they should solve the same task with the dataset they already knew from study one (cf. \autoref{sec:study_1_method_dataset}).
For both conditions, we provided instructions at two places: (1) in the tool itself (\colab: comments, \aisandbox: help-button), as a reminder, and (2) in the survey window.
Generally, the task was designed easy and broad to focus more on the general workflow experience and impression of the tools.
The task required from participants to
(1) interact with data (loading, deleting columns, visualization),
(2) train a model (configure splits and load a ML model),
(3) compare model performance with baseline model (F1-Score).
For example, in \colab, participants ran a provided cell to download the dataset, read the task in a text cell, and completed and executed a code cell to load the CSV into a pandas DataFrame (\autoref{fig:instruction-colab}).
The full description for both tools can be found in the supplementary materials.

\paragraph{Measurement}
\label{sec:study_2_method_measurement}
We employed UX and taskload assessments of \textit{AI Sandbox} and \textit{Colab} via UEQ-S \cite{schrepp_design_2017} (semantic differential scale from -3 to +3) and NASA-TLX \cite{hart_development_1988, hart_nasatask_2006} questionnaires (low/high endpoints on a 1-10 scale) completed after the use of each tool.
For acceptance we used the same single item ''I would use \mlguide for my academic/industrial machine learning workflow designs.'' as in study one (7-point Likert scale with endpoints: "fully disagree"/"fully agree").

\paragraph{Procedure}

\autoref{fig:procedure-2} shows a visual outline of the procedure.
The study consists of two subsequent phases in a randomly assigned and counterbalanced order where participants used both interactive programming environments and solved simple ML workflow design tasks (cf. \autoref{sec:study_2_method_task}). 
We asked for the solutions to the tasks in a quiz on the same page of the survey to keep participants engaged. 
Participants used \colab in the text-based programming condition; in the condition \textit{visual programming environment} they used \aisandbox. 
For \colab, we instructed participants to disable genAI-features, to keep the comparison to the base functionality.
In both conditions, we show a tutorial (cf. \autoref{app:instructions}) on how to interact with the tool and how to approach and solve the task (10min), before participants worked with the tool (15min).
Following every condition, we asked for participants taskload and UX perceptions.
In the \aisandbox condition, to better differentiate between three integrated tools \maratus (data editing), \mlkitchen (algorithm selection) and \mlflow (running, comparison), we additionally asked for UX and comments (positives, improvements, general) of the inidividual tool in the toolbox.  
After experiencing both environments, we asked participants for the acceptability of the \colab and \aisandbox so that they could make a direct comparison.
After that, both studies are finished and particpants filled out a final questionnaire (socio-demographics, general remarks).

\subsection{Results}

\paragraph{Workload (NASA-TLX)}
The overall workload was moderate for \aisandbox (M=4.74$\pm$1.83) and \colab (M=4.44$\pm$1.37).
The ratings show no significant difference between the two conditions ($\beta$=-0.23, t(14)=0.82, p=.42, d=0.20, CI$_{95}$[–0.36,0.84]), indicating that both tools require a similar level of mental effort, and participants’ cognitive load remained stable regardless of the system used. The only dimension where an effect is visible is the performance. An interaction between the programming environment and ML/programming-experience ($\beta_{\text{condition:xp}}$=-1.71, t(17)= -2.49, p=.024, d=0.61, CI$_{95}$[–3.18,0.24]):
participants with higher prior experience performed better in \colab{} than in \aisandbox{},
while less experienced participants showed no meaningful difference between tools.
In other words, \colab{} increasingly benefits more experienced users, whereas \aisandbox{} provides more uniform performance across experience levels.

\paragraph{Acceptance}
Acceptance ratings are generally high for both environments (M$_{\aisandbox}$=5.02$\pm$1.57; M$_{\colab}$=5.33$\pm$1.57), with no significant difference between them ($\beta$=-0.31, t(28)=0.55, p=.59, d=0.14, CI$_{95}$[–1.42,0.79]).
Overall, participants accept both systems well, perceiving them as equally appropriate and useful for their intended use in academia and practise.

\begin{figure*}[ht]
    \centering
    \begin{subfigure}[b]{0.3\textwidth}
        \includegraphics[width=\textwidth]{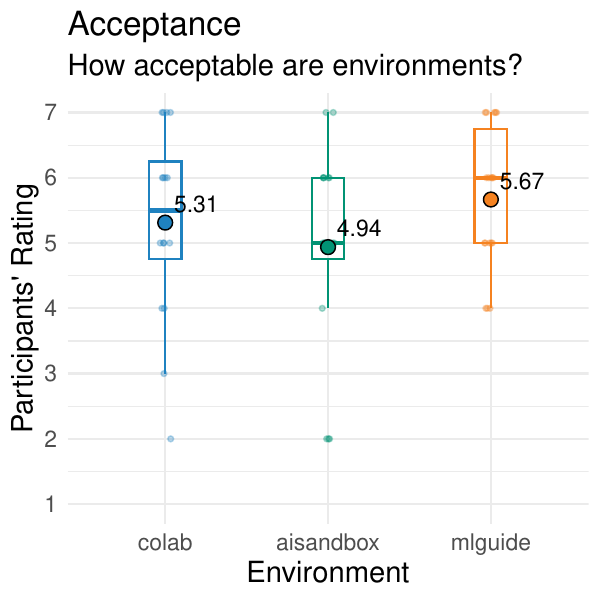}
        \caption{Acceptance ratings for study 1 (\mlguide{}) and study 2 (\colab{} and \aisandbox{}).}
        \label{fig:acceptance}
    \end{subfigure}
    \hspace{0.02\textwidth}
    \begin{subfigure}[b]{0.6\textwidth}
        \includegraphics[width=\textwidth]{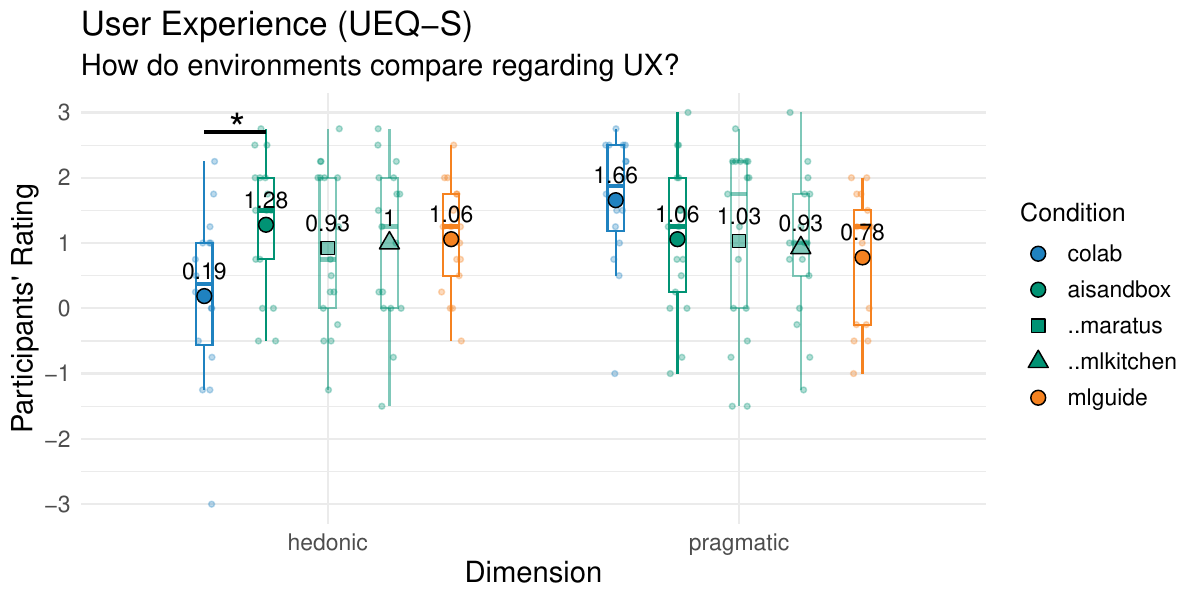}
        \caption{User Experience (UEQ-S) ratings for study 1 (\mlguide{}) and study 2 (\colab{} and \aisandbox{} with subsystems \maratus{}, \mlkitchen{}).}
        \label{fig:ueq}
    \end{subfigure}
    \begin{subfigure}[b]{0.62\textwidth}
        \includegraphics[width=\textwidth]{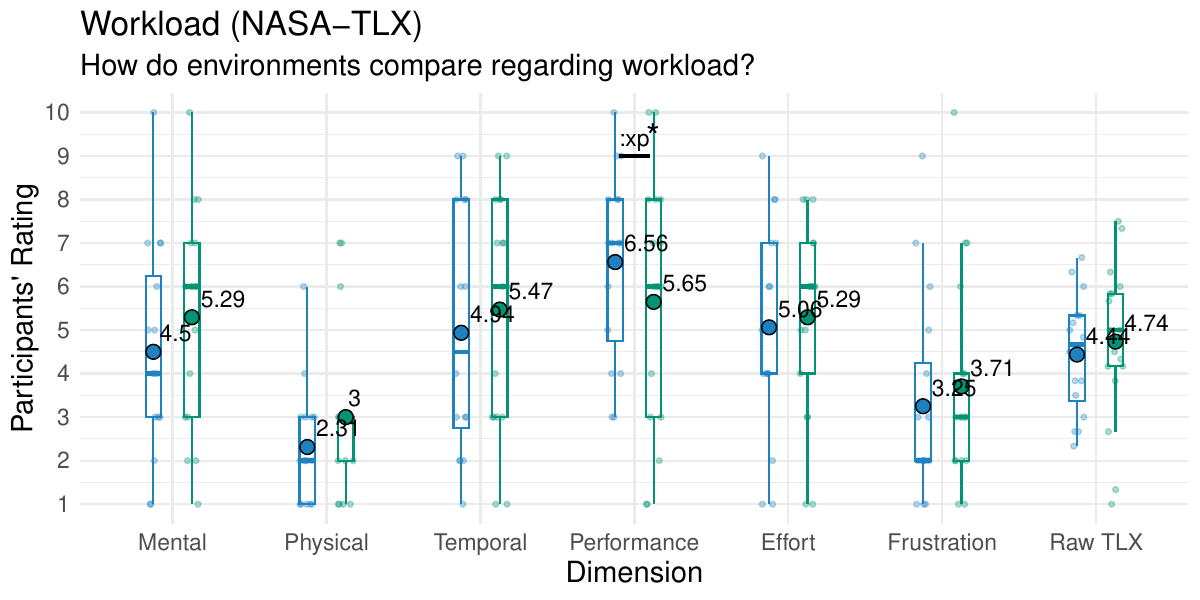}
        \caption{Workload}
        \label{fig:workload}
    \end{subfigure}
    \caption{Experience-related measures across the two studies. (a) Acceptance ratings for \mlguide in Study 1 and for \colab and \aisandbox in Study 2. (b) UEQ-S hedonic and pragmatic quality ratings. In Study 2, \aisandbox received higher hedonic ratings than \colab. (c) NASA-TLX workload ratings for the two Study 2 environments.}
    \label{fig:study2-experience-measures}
\end{figure*}

\paragraph{User Experience (UEQ-S)}
\autoref{fig:study2-experience-measures} summarizes the questionnaire results, including the UEQ-S ratings.
Regarding the \textit{hedonic quality}, ratings are rather high for \aisandbox (M=1.28$\pm$1.21) and neutral for \colab (M=0.16$\pm$1.21).
The difference is significant ($\beta$=1.12, t(29)=2.65, p=.013, d=0.90, CI$_{95}$[0.24,1.99]).
Participants rate both environments positively in terms of \textit{pragmatic quality}. There is a non-significant trend for higher values for \colab (M=1.66$\pm$0.99) compared to \aisandbox (M=1.06$\pm$1.15), yet showing comparable usability and clarity ratings ($\beta$=-0.59, t(15)=-1.89, p=.078, d=0.48, CI$_{95}$[–1.26,0.07]).
Participants also rated both \aisandbox components positively on the UEQ-S. For \maratus, the mean pragmatic score was $1.03\pm1.44$ and the mean hedonic score was $0.93\pm1.22$. For \mlkitchen, the mean pragmatic score was $0.93\pm1.08$ and the mean hedonic score was $1.00\pm1.22$.

Overall, participants rated both environments positively on pragmatic quality, while finding the visual environment more stimulating and engaging than \colab.

\section{Discussion}
\label{sec:discussion}
In this section, we discuss our results on the impact of structured guidance and programming environment on beginners ability to design and implement ML workflows, as well as the growing importance of ML education in many STEM contexts and the role of LLMs. 
Further, we point out implications for theory and practice and show the limitations of the study.

\subsection{Influence of Structured Guidance on Beginners' ML Workflow Conceptualization}
Our first research question (\textbf{RQ1}) asked: 
\textit{What is the influence of tool-based assistance (structured/unstructured) on the ML pipeline creators' conceptual understanding of ML pipelines?}

Results indicate that structured guidance through \mlguide improved participants’ high-level workflow design quality. 
After using the \mlguide, participants produced significantly better overall pipeline structures. Interestingly, this improvement is stronger for more experienced participants, suggesting that structured guidance may be especially useful when users already have enough background knowledge to interpret and integrate the recommendations effectively.
At the same time, \mlguide did not improve low-level methodological decision quality, such as selecting the correct data-splitting strategy, metrics, algorithms, or statistical tests. 
In \autoref{sec:design-challenge}, we argued that ML workflow design involves two intertwined challenges: users need to know \textit{what to do} (conceptual model) and \textit{how to do it} (implementation model), and assistance tools may help update these mental models differently.
Our findings suggest that \mlguide primarily externalizes and organizes evidence-backed recommendations, which makes best-practice structure easier to recognize, but still leaves the user with the burden of translating those recommendations into context-sensitive methodological judgments. 

These context-sensitive judgments were also time constrained in our study: more time spend with familiarization of the tool and deeper engagement with the task might yield different results in future studies. 
Confidence in the solution increased only slightly and non-significantly. 
Still, participants rated \mlguide positively in terms of pragmatic and hedonic UX qualities and reported high willingness to use it in academic or practical ML workflow design.

\mlguide acceptance and benefit is especially high for beginners in ML experimentation and, thus, relevant in domains such as medicine where rigour and transparency are critical. 
However, they should not be understood as substitutes for methodological education. 
Our study task was set in a clinical stroke-prediction scenario with requirements regarding false negatives, calibrated confidence, and interpretability. 
In such contexts, superficial pipeline correctness is not enough; beginners must understand the rationale behind decisions because methodological errors can directly undermine clinical trust, validity, and safety. Structured guidance could therefore be highly valuable in medical AI education as a first layer of support, but it should be embedded into teaching formats that also lead to critical appraisal, interpretability reasoning, and awareness of data-science pitfalls.

For designers of such systems, the key takeaway is to treat conceptual workflow support and methodological learning as \textit{two separate design targets}. 
Structured guidance can help beginners move closer to best practices, but it may create the impression that a structurally plausible workflow is also methodologically sound.
The difference between high- and low-level mental models show that we should distinguish between \textit{building a plausible pipeline} and \textit{being able to defend each modeling decision}.
A tool can improve pipeline completeness and correctness without necessarily improving principled reasoning. 
Designers should therefore add features such as (1) explanation-on-demand, (2) misconception-sensitive feedback, and/or (3) reflective checkpoints around known pitfalls like leakage, inappropriate metrics, and missing statistical comparison. 
The list of common pitfalls, modified dataset, and evaluation task provide a basis for designing and testing such interventions.

\subsection{Influence of Implementation Environment on Workload, UX, Acceptance}
Our second research question (\textbf{RQ2}) asked: 
\textit{What is the influence of interactive programming environments (text-based/visual) on the ML pipeline creators' task load, UX, and acceptance?} 
RAW-TLX results indicate that both environments \aisandbox and \colab pose comparable taskload on users. Participants find both environments acceptable for their academic or practical use. However, there are some in-depth findings where environments differ in task and experience-oriented aspects.
We found a significant interaction of general ML/programming experience with performance, i.e., the text-based approach felt more effective for experienced users compared to the visual approach. 
Additionally, there is a non-significant trend in the UEQ-S pragmatic scale dimension. 
On the other hand, \aisandbox' visual programming environment got significantly higher hedonic UEQ-S ratings across all users and expertise levels.

More experienced participants rated their performance more positively in the familiar text-based environment, while still valuing the visual environment for its hedonic qualities.
For less experienced participants, perceived performance was comparable across environments. The visual environment may make workflow design feel more approachable.
A visual workflow representation can make pipeline structure more tangible and less intimidating: A visual graph may reduce the mismatches between users’ mental models and the system’s representation (cf. \autoref{sec:design-challenge}), even if it does not reduce the complexity of the task. 
This is especially plausible in ML workflow design, where cognitive demands come not only from syntax and coding but from understanding dependencies among different steps in preprocessing, modeling, and evaluation decisions. 
Without automatically simplifying the underlying reasoning challenge, the visual environment may improve accessibility and engagement, and, thus, can influence beginners willingness to engage with the complex tasks of ML workflow design. 
Nevertheless, visual programming environments should be understood as supportive representations rather than shortcuts around the need to learn core ML concepts.

Designers of visual ML environments can use the interface to expose dependencies, explain the consequences of choices, and let users inspect the effects of each workflow step.
For certain domains, such as medical education, the visual programming environment might be especially attractive. 
Medical learners often need to understand the logic of an ML experiment without necessarily becoming full programmers. A no-code, visually explicit system may therefore help them participate in model design discussions, reason about data-processing and evaluation steps, and engage more confidently with AI-based decision-support systems. 
This is particularly relevant in healthcare, where transparency, interpretability, and awareness of methodological pitfalls are essential for trust and responsible adoption. 
A visual environment could thus serve as a bridge between clinical reasoning and computational workflow thinking.

\subsection{Integration of Conceptual Help in ML Implementation Tools: Interaction Model}
The two studies suggest combining structured conceptual guidance with an accessible implementation environment. \autoref{fig:problem} illustrates the cycle this integration would support: when users encounter a problem during implementation, they seek guidance, select an approach, and return to the workflow to apply it. Bringing assistance into the workflow editor would let users inspect the evidence behind a recommendation, compare alternatives, and see how their choice changes the pipeline.

Our toolbox illustrates how these roles can be combined. \maratus makes data preparation visible and editable, \mlguide provides evidence-backed methodological guidance, \mlkitchen turns design choices into an executable configuration, and \mlflow records runs for comparison. Future systems could automate selected steps according to users' experience while keeping the rationale, alternatives, and consequences visible. They could also ask users to review and justify consequential choices before execution. These mechanisms give users practice in methodological judgement and make it easier to exercise oversight when automation can otherwise contribute to skill loss or overreliance \cite{bainbridge1983ironies,parasuraman1997automation}.

\subsection{Limitations}

The sample was small and predominantly technical, although the broader motivation includes domain experts such as clinicians. The findings therefore provide preliminary evidence about how beginner-oriented tools can support conceptual and implementation skills.

\paragraph{Study 1}

The absence of low-level improvements is also theoretically meaningful. It suggests that methodological depth in ML workflow design may require more than recommendation lists or rankings. For example, users may need richer explanations, worked examples, counterexamples, or interactive reflection prompts that explicitly connect recommendations to task constraints such as class imbalance, explainability, or evaluation validity. This interpretation also fits the methodological challenges identified in our design rationale, where pitfalls such as leakage, unsuitable metrics, or inappropriate validation strategies require context-sensitive reasoning rather than only awareness of workflow stages.

Confidence in the solution increased only slightly and non-significantly. One explanation is that structured guidance may have made participants more aware of the complexity of the task and of possible shortcomings in their initial designs. At the same time, participants rated \mlguide positively in both pragmatic and hedonic UX qualities and reported high willingness to use it in academic or practical ML workflow design. Taken together, this suggests that users value structured conceptual support even when it does not immediately translate into stronger methodological quiz performance.

\paragraph{Study 2}
Future work should look into the long-term impact of continuous use to determine whether workload and UX change after the initial introduction phase. 

\section{Conclusion}

This paper examined two complementary ways of supporting beginner ML experimentation: structured, evidence-backed guidance for conceptual workflow design, and a visual no-code environment for workflow implementation. Across two within-subject studies, the results point to an important distinction. Structured guidance helped participants produce more plausible high-level workflow structures, but it did not measurably improve low-level methodological reasoning. The visual no-code environment improved hedonic user experience relative to a notebook-based baseline, while perceived workload and acceptance remained broadly similar. Taken together, these findings suggest that different forms of support scaffold different layers of ML workflow competence.

For interactive intelligent systems, this has a clear design implication. Beginner-facing ML tools should not be evaluated as though usability, automation, and methodological quality were interchangeable outcomes. A system may improve workflow composition without improving methodological judgement, or improve user experience without reducing conceptual difficulty. We therefore argue for mixed-initiative ML tools that combine structural guidance, inspectable recommendations, and opportunities for reflection about local methodological choices. At the same time, our results should be interpreted with caution given the fixed-order design of Study~1, the small and predominantly technical sample, and measurement limitations. Rather than demonstrating broad learning transfer, the present work identifies a more specific design opportunity: to connect conceptual scaffolding and implementation support more tightly, so that future systems can help users not only build workflows that look plausible, but also justify why those workflows are methodologically appropriate.

\begin{acks}
We would like to thank the entire project team for their fruitful discussions and invaluable support, particularly Rieke Alpers, Johannes Bunk, Lucas Rebscher, and Elham Taghizadeh.

This work was funded based on a resolution of the German parliament by the German Federal Ministry of Health (BMG, Project KI-FDZ, Funding Code: 2521DAT01B). The funding bodies played no role in the design of the study and collection, analysis, and interpretation of data and in writing the manuscript.
\end{acks}

\bibliographystyle{ACM-Reference-Format}
\bibliography{bibliography/base, bibliography/references, bibliography/zotero}

\appendix
\section{Participants' Details}
\label{app:participants}
\begingroup
\small
\setlength{\tabcolsep}{2pt}
\begin{tabular}{@{}p{0.07\linewidth}p{0.10\linewidth}p{0.15\linewidth}p{0.45\linewidth}p{0.14\linewidth}@{}}
\toprule
\textbf{ID} & \textbf{Age} & \textbf{Gender} & \textbf{Field} & \textbf{Location} \\
\midrule
P1 & 21 & female & Computer Science & on-site \\
P2 & 27 & male & Digital Media \& Software Dev & online \\
P3 & 23 & male & AI \& Intelligent Systems (MSc) & online \\
P4 & 24 & male & Digital Media (MSc) & online \\
P5 & 22 & male & AI \& Intelligent Systems (MSc) & online \\
P6 & 30 & male & Control, Microsystems \& Microelectronics (MSc) & online \\
P7 & 24 & male & Artificial Intelligence & online \\
P8 & 27 & male & Control, Microsystems \& Microelectronics (MSc) & online \\
P9 & 23 & male & AI \& Intelligent Systems (MSc) & on-site \\
P10 & 27 & male & Process \& Energy Engineering & on-site \\
P11 & 24 & male & Data Science & on-site \\
P12 & NA & NA & NA & online \\
P13 & 23 & male & Electronics \& Telecommunications & on-site \\
P14 & 28 & male & Control, Microsystems \& Microelectronics (MSc) & on-site \\
P15 & 31 & non-binary & Data Engineering & online \\
P16 & 22 & female & Student (Undeclared) & on-site \\
P17 & NA & NA & NA & on-site \\
P18 & 27 & female & International Relations & on-site \\
\bottomrule
\end{tabular}
\par\noindent Age is in years. NA = not disclosed by participant.
\par\endgroup

\section{Study 2 – Tool Instruction Examples}
\label{app:instructions}
\begin{figure}[h]
    \centering
    \includegraphics[width=\linewidth]{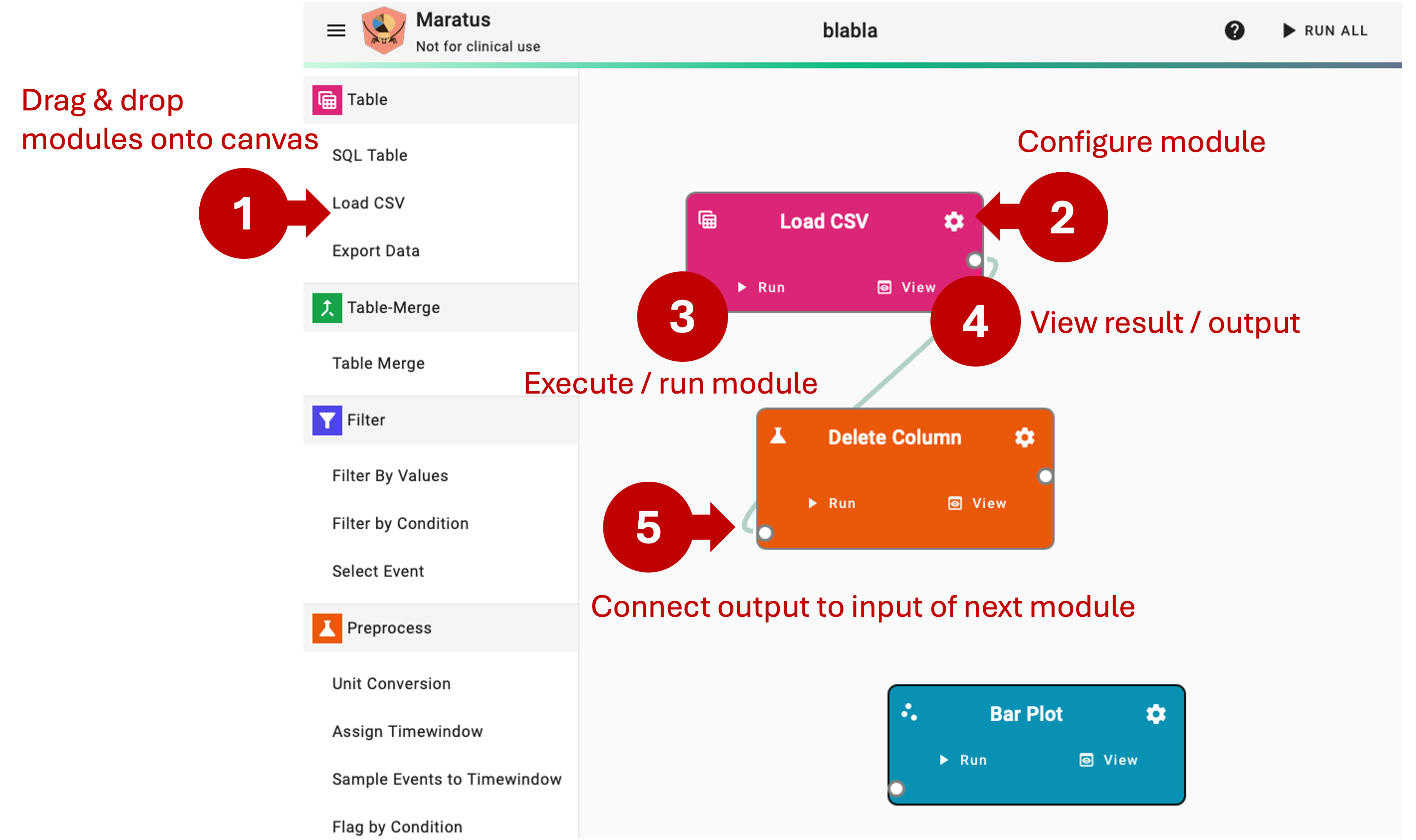}
    \caption{Example instructions on how to use the tool \textit{maratus}}
    \label{fig:instruction}
\end{figure}
\begin{figure}[ht]
    \centering
    \includegraphics[width=\linewidth]{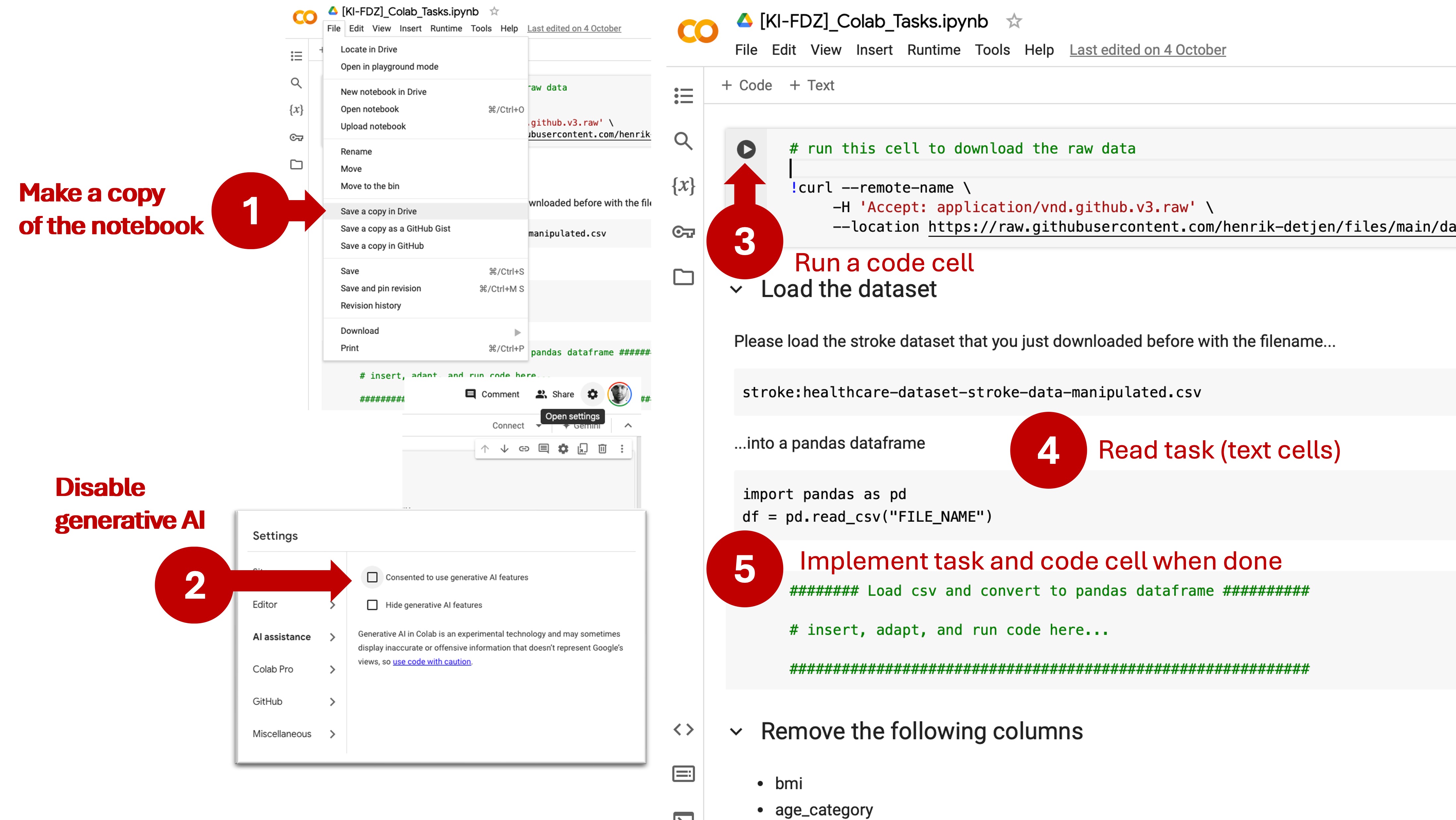}
    \caption{Example instructions for the Google Colab condition in Study 2. Participants copied the notebook, disabled generative AI, ran the supplied download cell, and completed the task in the indicated code cell.}
    \label{fig:instruction-colab}
\end{figure}

\section{Task 1.0 -- Overall Briefing and Dataset}
\label{app:task1.0}

\subsection{Overall Briefing}
\label{app:briefing}
You are working as a data scientist in a hospital. You have been provided with data from a physician working at a stroke unit. The physician asks you to develop a machine learning (ML) model that is predicting stroke events from patient-related clinical and demographic data. The model, if it works well, is intended to be used in clinical practice for early detection of patients with a high risk of stroke in order to initiate important changes in lifestyle and / or medication as early as possible.
Familiarize yourself with the data. Look at all the columns, and learn about their properties and distributions.
You are asked to do two things:
\begin{enumerate}
    \item You need to design an ML workflow for solving the model development task described above.
    \item You will be asked which methods you would choose for some selected workflow steps.
\end{enumerate}
Here is a high-level overview of what you need to do, providing you with hints on how to design the workflow and on methodological requirements:
\begin{enumerate}
    \item Prepare the data so that it can be used for the model training and evaluation process.
    \item Select and train a suitable ML model to classify unseen persons’ data. Try several state of the art ML models.
    \item Evaluate the performance of your model with suitable metrics.
    \begin{enumerate}
        \item The physician tells you that false negatives (FNs) are more severe than false positives (FPs) because patients with a high risk of stroke that the model misses (FNs) might miss important treatments that help them stay healthy. Conversely, if a patient has a low risk of stroke but the model classifies the patient as high risk (FP), this is also bad because of a waste of time and resources for patients and healthcare staff but less severe than an FN.
        \item Clinicians demand that the model outputs a reliable measure of how sure it is about its predictions.
    \end{enumerate}
    \item Compare your result with the logistic regression model that is currently used for stroke prediction in practice. The two models should be compared on an independent test set. In the end, you want to be able to confidently demonstrate that your new model is better than the current model. Quality management demands that you provide rigorous statistical tests assessing the performance difference to the current model.
    \item Physicians using the model should be able to comprehend which input variables are important for the model predictions so that they can interpret and understand prediction results. 
\end{enumerate}
You can take notes and use any kind of help you would normally use. Here are a few tools that might be helpful:
\begin{itemize}
    \item Chat-GPT (or similar) 
    \item StackExchange (or similar) 
\end{itemize}

\subsection{Dataset with Modifications}
\label{app:dataset}
\begin{figure*}[h]
    \centering
    \includegraphics[width=\linewidth]{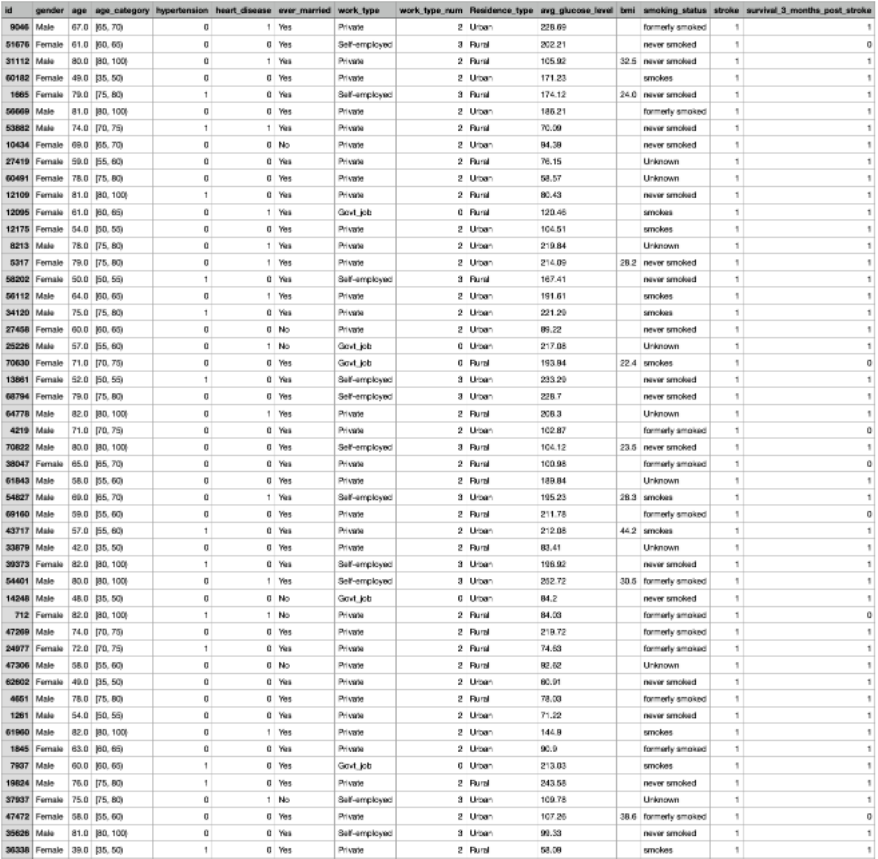}
    \caption{Excerpt of the modified stroke prediction dataset. The first row (ID~9046) illustrates an added age band, an integer employment code, a blank BMI, and a synthetic three-month survival value. The excerpt contains stroke-positive records.}
    \label{fig:dataset_excerpt}
\end{figure*}

The source was the public Stroke Prediction Dataset published by \textit{fedesoriano} on Kaggle \cite{dataset-orginal}.
We applied the manipulations described in \autoref{sec:study_1_method_dataset}.
\autoref{fig:dataset_excerpt} shows how several of the manipulations appear in the data:
\begin{itemize}
    \item \textbf{Age bands and column order.} For ID~9046, \texttt{age} is 67.0 and the adjacent added \texttt{age\_category} is \texttt{[65, 70)}. The excerpt also places the integer employment code next to its source and survival after the stroke outcome.
    \item \textbf{Redundant employment code.} For ID~9046, \texttt{work\_type} is \texttt{Private} and the added \texttt{work\_type\_num} is 2. ID~51676 shows \texttt{Self-employed} encoded as 3.
    \item \textbf{Synthetic survival and leakage.} IDs~9046 and~51676 both have \texttt{stroke}=1 and show synthetic three-month survival values of 1 and 0, respectively. The survival field is blank for stroke-negative records in the full dataset. The excerpt shows only stroke-positive records, so those blanks are not visible here.
    \item \textbf{Outcome-dependent BMI missingness.} ID~9046 has \texttt{stroke}=1 and a blank \texttt{bmi}, compared with 36.6 in the original CSV. The excerpt cannot show the imputation of missing BMI for stroke-negative records because it contains none.
    \item \textbf{Duplicated rows.} ID~70822 is visible once in the excerpt and has an exact copy in the full modified CSV. No duplicate pair is visible in this excerpt.
\end{itemize}
The supplementary materials include the full modified dataset and the manipulation script.

\section{Task 1.1 -- Step-wise Workflow Design with Distractors}
\label{app:task1.1}
\subsection{Instructions}
You are offered a list of potential workflow steps for the task of developing and evaluating the stroke prediction model. Create a workflow that is suitable for this task by selecting the workflow steps that need to be done from the list of workflow steps that you find below, and bring the selected workflow steps into a valid order (some steps need to be done before certain other steps).

\subsection{Steps \& Rating}

The following workflow steps are considered wrong answers (they shouldn't be included in the workflow):

\begin{itemize}
    \item Trend Detection
    \item Backtesting
    \item Data Collection
    \item Model Monitoring
    \item Anomaly Detection
    \item Data Synthetization
    \item Data Homogenization
\end{itemize}
The following workflow steps are considered not mandatory, but also not wrong (if included in the workflow, we check if the ordering of workflow steps makes sense. If it does, the workflow is considered valid):

\begin{itemize}
    \item Ensembling
    \item Model Deployment
    \item Oversampling
    \item Outlier Replacement
    \item Feature Selection
    \item Dimensionality Reduction
    \item Feature Engineering
    \item Model Calibration
\end{itemize}
The following workflow steps are considered correct. They are listed in a correct ordering, but there are also other correct orderings, see "Expected solution (ordering)":

\begin{itemize}
    \item Data Preparation
    \item Data Splitting
    \item Categorical Feature Encoding
    \item Missing Data Handling
    \item Metric Selection
    \item Algorithm Preselection
    \item Hyperparameter Tuning
    \item Model Training
    \item Model Selection
    \item Model Evaluation
    \item Model Interpretation
\end{itemize}
Study participants will be presented with a randomly ordered list of the workflow steps listed above. They need to select the correct workflow steps and bring them into a correct order.

\subsubsection{Expected solution (ordering)}

In order to evaluate the solutions of participants, we don't only check if correct workflow steps have been selected, but also if they have been selected in the correct order. The following rules for ordering should be satisfied by a correct solution. This list is not meant to be exhaustive, but these are the conditions that we want to check in the evaluation. "Step A < Step B" means "Step A should occur prior to Step B in the workflow".

For mandatory steps:
\begin{itemize}
    \item Model Training < Model Selection
    \item Model Selection < Model Evaluation
    \item Hyperparameter Tuning < Model Selection
    \item Algorithm Preselection < Hyperparameter Tuning
    \item Algorithm Preselection < Model Training
    \item Metric Selection (Discrimination) < Model Selection
    \item Metric Selection (Calibration) < Model Selection
    \item Data Splitting < Hyperparameter Tuning
    \item Data Splitting < Model Training
    \item Data Preparation < Data Splitting
    \item Missing Data Handling < Model Training
    \item Missing Data Handling < Hyperparameter Tuning
    \item Categorical Feature Encoding < Model Training
    \item Categorical Feature Encoding < Hyperparameter Tuning
    \item Model Training < Model Interpretation
\end{itemize}

For optional steps:

\begin{itemize}
    \item Model Training < Ensembling
    \item Model Training < Stacking
    \item Model Training < Model Deployment
    \item Data Splitting < Oversampling
    \item Outlier Replacement < Model Training
    \item Model Training < Decision Threshold Tuning
    \item Data Splitting < Feature Selection
    \item Data Splitting < Dimensionality Reduction
    \item Feature Engineering < Data Splitting
    \item Metric Selection (Fairness) < Model Selection
    \item Model Training < Model Calibration
\end{itemize}

\subsubsection{Evaluation Strategy}

\begin{itemize}
    \item +2 for each correctly selected workflow step
    \item -1 for each missing correct workflow step
    \item -1 for each violation of ordering rules
    \item -1 for each wrongly selected workflow step
    \item 0 for each optional workflow step
\end{itemize}

\section{Task 1.2 -- Methodological Questions for Specific Workflow Steps}
\label{app:task1.2}

We have picked 5 workflow steps for which we ask participants in-depth methodological questions:

\begin{enumerate}
    \item Data Preprocessing / Data Cleaning (2 questions on this topic)
    \item Data Splitting
    \item Metric Selection
    \item Algorithm Preselection
    \item Statistical Analysis
\end{enumerate}

\subsection{Task 1.2.1 -- Data Cleaning}

The data needs to be "clean" before you start model development. Which of the following issues can you identify in the stroke prediction dataset?

\noindent[select multiple]
\begin{enumerate}
    \item[\XBox] Multiple tables: Datasets from different tables need to be merged to create a single tabular dataset,
    \item[\CheckedBox] Redundant features: Features that provide similar information or contain duplicated information.
    \item[\CheckedBox] Duplicate rows: Rows that are identical and result in redundant data points.
    \item[\XBox] Inconsistent labeling: Variations in labeling for categorical data that can cause inconsistencies.
    \item[\CheckedBox] Missing values: Data points that are missing in the dataset.
    \item[\XBox] Raw text data: Raw text data that requires cleaning and preprocessing. (e.g., removal of stop words, stemming).
    \item[\XBox] Implausible values: Anomalous data points that are unrealistic (e.g., an age of 200).
    \item[\XBox] Structural errors: Typos and inconsistent capitalization in the dataset.
    \item[\XBox] Inconsistent units: Variations in units of measurement across the dataset.
    \item[\CheckedBox] Leakage-prone features: Features that include future information or information not available at prediction time.
    \item[\CheckedBox]  Irrelevant features: Features that do not contribute to the predictive power of the model.
\end{enumerate}

\subsection{Task 1.2.2 -- Feature Engineering}

\subsubsection{Task Description}

Depending on their types and distributions, as well as on domain knowledge, each feature might need to be handled differently during data preprocessing.
For each feature in the stroke data set, decide which operations need to be done in order to prepare the data for training an ML model.
Hint: For most of the features, only one preprocessing step is needed. There are also features for which no preprocessing is necessary.

Potential operations [assign 0-1 to every feature]:
\begin{itemize}
    \item Remove feature
    \item One-hot encoding
    \item Ordinal encoding
    \item Target encoding
    \item Missing values imputation
    \item Handle low frequency categories
    \item Binning
    \item Min-max scaling
    \item Standardization
    \item Log transformation
    \item Outlier replacement
\end{itemize}

\subsubsection{Expected solution}
\autoref{tab:task-2-2} shows the expected solution.
\begin{table*}[t]
    \centering
    \begin{tabular}{lll}
        \toprule
        \textbf{Feature} & \textbf{Mandatory Operation(s)} & \textbf{Optional Operation(s)} \\
        \midrule
        id & Remove column (no information) & None \\
        gender & Handle low frequency categories & None \\
        age & None & Min-max scaling or Standardization \\
        age\_category & Remove column (binned version of age) & None \\
        hypertension & None & None \\
        heart\_disease & None & None \\
        ever\_married & None & One-hot encoding or Target encoding \\
        work\_type & One-hot encoding or Target encoding & None \\
        work\_type\_num & Remove column & None \\
        residence\_type & None & One-hot encoding or Target encoding \\
        avg\_glucose\_level & None & Min-max scaling or Standardization \\
        bmi & Remove feature & None \\
        smoking\_status & None & One-hot encoding or Target encoding \\
        stroke & None & None \\
        survival\_3\_months\_post\_stroke & Remove column & None \\
        \bottomrule
    \end{tabular}
   \caption{All mandatory and optional transformations for the feature engineering step.}
   \label{tab:task-2-2}
\end{table*}

\subsubsection{Evaluation Strategy}

For each column:
\begin{itemize}
    \item +1 point if all mandatory transformations are selected and no transformation outside the mandatory or optional sets is selected.
    \item +0 points otherwise
\end{itemize}

\subsection{Task 1.2.3 -- Data Splitting}

\subsubsection{Task Description and Evaluation}

Based on the task description and your knowledge on how to train and evaluate ML models, which data splitting strategy would you choose? 

\noindent[select one]
\begin{itemize}
    \item Train-Test (2 Points)
    \item Train-Validation-Test (5 points)
    \item Cross-Validation with independent test set (6 points)
    \item Bootstrapping with independent test set (6 points)
    \item Nested Cross-Validation (4 points)
    \item Bootstrapping (0 points)
    \item No Split (0 points)
    \item Cross-Validation (0 points)
    \item Monte Carlo Cross-Validation (0 points)
\end{itemize}

\subsubsection{Evaluation Strategy}

Reasoning for the assigned points above. 
Assign points for each of the following properties if it is fulfilled for the chosen data splitting strategy:

\begin{itemize}
    \item We need to have both a validation set (for model selection) and a test set (for model evaluation) (+3 points)
    \item Choose a splitting strategy that does multiple resampling on the validation set for more robust model selection (as compute resources don't matter). (+1 point)
    \item Choose a single test set so that a specific model can be evaluated because we want a validated model for use in production (+2 points)
\end{itemize}

\subsection{Task 1.2.4 -- Metric Selection}

\subsubsection{Task Description and Evaluation}

Based on the task description and your knowledge on Ml model evaluation, which primary metric would you choose to evaluate the discriminative performance of your model?

\noindent[select one]
\begin{itemize}
    \item Weighted mean of sensitivity and specificity (8 points)
    \item Accuracy (3 points)
    \item Balanced Accuracy (6 points)
    \item Cost-sensitive Accuracy (8 points)
    \item Sensitivity (5 points)
    \item Specificity (4 points)
    \item Precision (4 points)
    \item F1-Score (6 points)
    \item Fbeta-Score (8 points)
    \item AUROC (6 points)
    \item AUPRC (6 points)
    \item Phi coefficient (6 points)
\end{itemize}

\subsubsection{Evaluation Strategy}

Assign points for each of the following properties if it is fulfilled for the chosen metric:

\begin{enumerate}
    \item Suitable for imbalanced data (+3 points)
    \item Considers FPs (+1 point)
    \item Considers FNs (+2 points)
    \item Is cost sensitive (+2 points)
\end{enumerate}

\subsection{Task 1.2.5 -- Algorithm Preselection}

\subsubsection{Task Description}

Based on the task description and your knowledge on Ml algorithms, which algorithms would you experiment with and compare for model training in order to come up with a good model? Multiple options are allowed as you can experiment with several algorithms before you select the one that works best.

Correct answers:
\begin{itemize}
    \item Support Vector Machine (Group: SVM)
    \item Multilayer Perceptron (Group: Neural Networks)
    \item k-NN (Group: Nearest Neighbors)
    \item Decision Tree (Group: Decision Trees)
    \item Random Forest (Group: Decision Tree Ensembles)
    \item XGBoost (Group: Boosting via Decision Trees)
    \item LightGBM (Group: Boosting via Decision Trees)
    \item Naive Bayes (Group: Naive Bayes)
    \item Logistic Regression (Group: Linear Models)
    \item Stochastic Gradient Descent (Group: Linear Models)
\end{itemize}

Distractors:
\begin{itemize}
    \item Linear Regression (Regression algorithm)
    \item Transformer (NLP algorithm)
    \item LSTM (Time series algorithm)
    \item Convolutional Neural Network (Image data algorithm)
\end{itemize}

\subsubsection{Evaluation Strategy}

\begin{itemize}
    \item +3 points if at least two correct algorithms are selected
    \item -2 points for every wrong selection
    \item +1 point for every inherently interpretable algorithm selected (capped at 3 points)
    \item +1 point for every group from which at least one algorithm is selected (capped at 3 points)
\end{itemize}

\subsection{Task 1.2.6 -- Statistical Analysis}

How would you evaluate your ML model in comparison to the benchmark model?

\noindent[select one]
\begin{itemize}
    \item[\XBox] Difference of Point Estimates: Compute the difference of the point estimates of the primary performance metric between the new model and the benchmark model 
    \item[\XBox] Paired t-test: Conduct a t-test on the differences of performance estimates between the new model and the benchmark model on the test sets: Compare the difference in error rates of the two models on the test set
    \item[\XBox] Normal approximation: Assume that the difference in performance is approximately normally distributed in order to conduct appropriate statistical tests
    \item[\XBox] McNemar’s test: Compare the difference in error rates of the two models on the test set
    \item[\CheckedBox] Bootstrap test set: Bootstrap the test set and compute differences of performance estimates for each bootstrapped sample in order to construct confidence intervals
\end{itemize}

\end{document}